\documentclass[preprint,12pt]{elsarticle}

\usepackage{lineno,hyperref}
\modulolinenumbers[5]

\journal{Combustion and Flame}

\usepackage{amssymb,amsmath}
\usepackage{graphicx}

\usepackage{comment}

\usepackage{textgreek}
\newcommand{\gmu}{\textmugreek}

\usepackage{color,soul}

\usepackage{cleveref}
\crefformat{section}{\S#2#1#3} 
\crefformat{subsection}{\S#2#1#3}
\crefformat{subsubsection}{\S#2#1#3}

\begin{document}

\begin{frontmatter}
	
	\title{Investigation of Turbulent Mixing and Local Reaction Rates on Deflagration to Detonation Transition in Methane-Oxygen}

	\author[uvic,uo]{Brian Maxwell\corref{cor1}}
	\ead{brianmaxwell@uvic.ca}
	\author[shell]{Andrzej Pekalski}
	\author[uo]{Matei Radulescu}

	\cortext[cor1]{Corresponding Author}
	\address[uvic]{Institute for Integrated Energy Systems, University of Victoria, PO Box 1700 STN CSC, Victoria, Canada, V8W 2Y2}
	\address[shell]{Shell Global Solutions, Brabazon House, Concord Business Park, Threapwood Road, \\M22 0RR Manchester, United Kingdom}
	\address[uo]{Department of Mechanical Engineering, University of Ottawa, 161 Louis Pasteur, Ottawa, Canada, K1N 6N5}

	
	
	
	
	\begin{abstract}
		In the current study, the influence of turbulent mixing and local reaction rates on deflagration to detonation transition (DDT) was investigated using a state-of-the-art large eddy simulation (LES) strategy.  Specifically, detonation attenuation by a porous medium, and the subsequent re-initiation for methane-oxygen, a moderately unstable mixture, was considered.  The purpose of the investigation was to validate the numerical strategy with previous experimental observations, and to determine what specific roles turbulent mixing and shock compression have on flame acceleration during the final stages of DDT. The modelling procedure adopted was a grid-within-a-grid approach: The compressible linear eddy model for large eddy simulation (CLEM-LES).   It was found that average turbulent fluctuations greater than the laminar flame speed were required in order to maintain wave velocities above the Chapman-Jouguet (CJ)-deflagration velocity threshold, a precursor requirement for DDT to occur.  It was also found that sufficient turbulent burning on the flame surface was required in order to drive pressure waves to sufficiently strengthen the leading shock wave, locally, in order to trigger auto-ignition hot spots in the wave front.  These local explosion events, which were found to burn out through turbulent surface reactions, drive transverse pressure waves outward.  Upon subsequent shock reflections or interactions of the transverse waves, new local explosion events occurred, which further strengthened the adjacent leading shock wave above the CJ-detonation speed.  Eventually, through this process, the wave sustained the CJ-detonation speed, on average, through the cyclic mechanism of local explosion events followed by turbulent surface reactions.  Finally, combustion of the flame acceleration process was found to lie predominantly within the thin-reaction zones regime.
	\end{abstract}
	

\end{frontmatter}



\section{Introduction}

\begin{figure}[ht]
\centering
\includegraphics[scale=0.7]{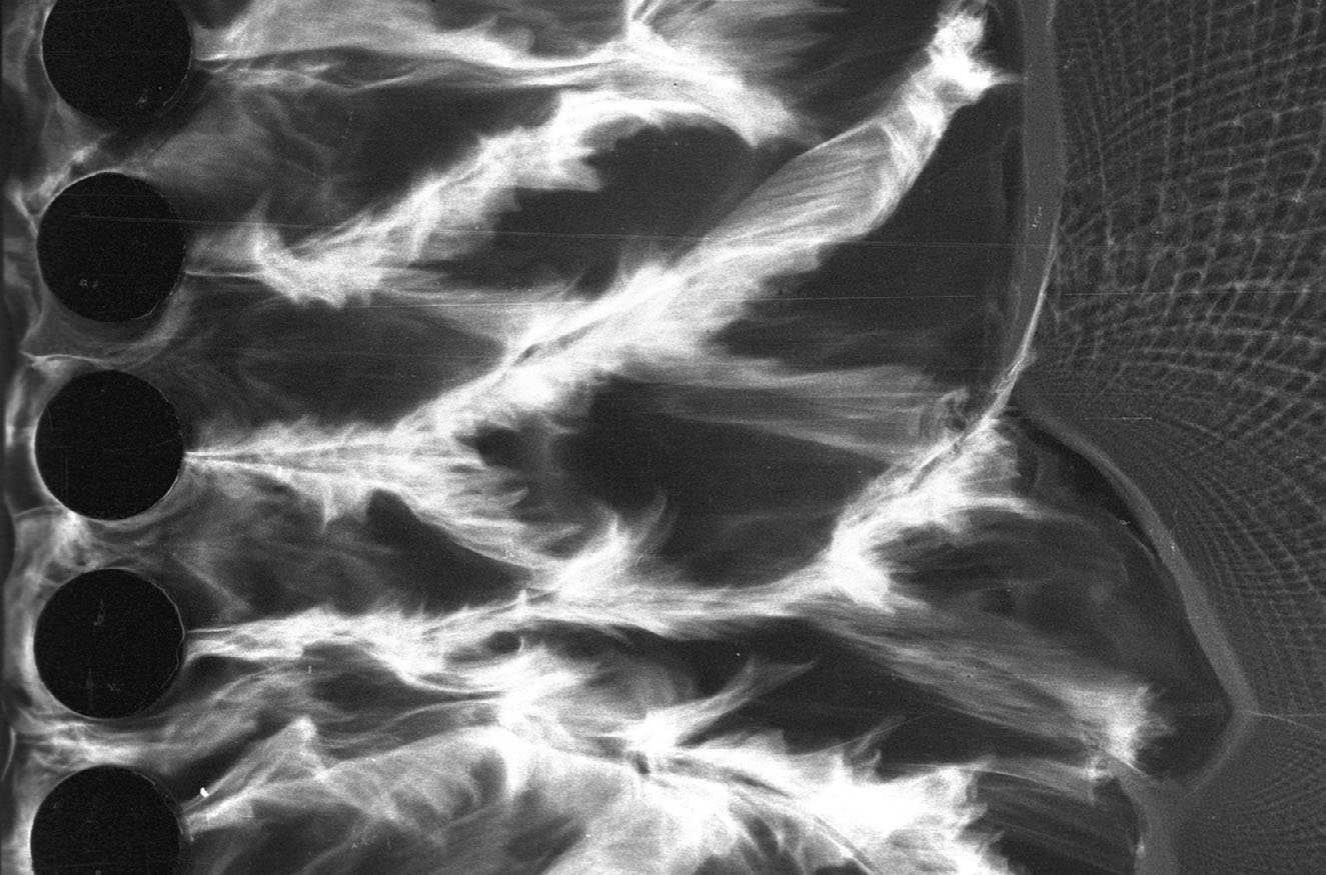}
\caption{Open shutter photograph showing DDT of $\textrm{C}_2\textrm{H}_2+\textrm{O}_2$ at $\hat{p}_o=5$ kPa, following detonation interaction with a porous medium \cite{Radulescu2011}.}
\label{fig.Radulescu2011_exp}
\end{figure}

In the current study, the influence of \emph{turbulent mixing} rates and local reaction rates on deflagration to detonation transition (DDT) in methane-oxygen was investigated, using a state-of-the-art large eddy simulation (LES) strategy.  Particular focus on understanding the physical mechanisms which contribute to DDT has been an active area of research in the wake of a number of recent accidents involving compressed gases.  One well publicized incident, the Buncefield explosion \cite{BMIB2008}, highlighted DDT as a major physical fundamental problem, which is not yet clearly understood.  In 2005, release of hydrocarbon vapours at the Buncefield oil storage depot, England, triggered an explosion much more powerful than anyone had anticipated probable.  It has been speculated that detonation occurred as the ensuing reactive wave propagated along a row of trees \cite{Johnson2010}.  It is believed that the presence of obstacles allowed for increased turbulent mixing in the reaction zone and thus gave rise to enhanced combustion rates.  In order to understand how detonations, or powerful explosions occur on the industrial scale, it is necessary to understand the underlying physics of how flames accelerate in compressed fluids and transition to detonation, and the mechanisms by which detonations can sustain propagation.  Also, advancement of detonation based engines, for supersonic propulsion, rely on the ability to predict and control DDT events \cite{Pandey2016}.  In order to advance the state of knowledge and fundamental understanding of DDT, the scenario investigated here corresponded to recent physical experiments \cite{Maley2015,Ahmed2016,Saif2016}.  More specifically, this investigation followed the procedure of Radulescu and Maxwell \cite{Radulescu2011} by considering the re-ignition of fully quenched detonations, which followed detonation interaction with a porous medium, as shown in Fig.\ \ref{fig.Radulescu2011_exp}.  This type of DDT has also been examined experimentally for detonation interactions with perforated plates \cite{Zhu2007}, or a series of obstacles, or blockages \cite{Lyamin1991,Teodorczyk1988}.

Currently, the quenching process of detonations is well understood.  As a detonation wave diffracts around an object, the sudden change in area causes volumetric expansion of the gas behind the leading shock wave.  Eventually, the detonation can become quenched when local cooling due to this expansion overcomes local heating due to chemical reactions \cite{Arienti2005,Eckett2000,Radulescu2010}.  The result is a de-coupling between the leading shock wave and reaction zone, as observed experimentally \cite{Soloukhin1969,Lundstrom1969}.  The quenched detonation is thus a shock wave followed by a turbulent deflagration.  This type of quenching has also been observed numerically for rapidly expanding diffusion layers applied to jet ignition of hydrogen \cite{Maxwell2011a}.  The re-initiation process, however, is not so clear.  To date, it has been found that re-initiation of the attenuated detonation wave occurs through amplification of the incident shock strength resulting from shock reflections or triple point collisions \cite{Radulescu2005,Radulescu2011}.  In some cases, several shock reflections were required to accelerate the leading shock wave sufficiently in order to re-initiate the detonation.  At each shock reflection, or triple point collision, the incident shock accelerates due to increased reaction rates in the un-burned gases behind the incident shock.  In similar experiments, which examined {\emph{quasi-detonation}} propagation in {porous media} \cite{Makris1993b,Makris1995,Slungaard2003}, it has been shown that {a wave} can be sustained below the steady Chapman-Jouguet (CJ) detonation velocity \cite{Fickett1979}.  Due to the velocity deficit, it is believed that \emph{adiabatic compression} alone, from shock interactions, cannot provide the necessary ignition to sustain a detonation wave.  Thus, it remains unclear whether adiabatic shock compression or turbulent mixing is the dominant mechanism that drives, or initiates, the detonation. 

{In order to investigate the role of shock interactions on detonation re-initiation of irregular mixtures, following wave interaction with porous media, a numerical strategy based on the Euler formulation was previously adopted} \cite{Radulescu2011}.  {Unfortunately, the numerical strategy, which did not address turbulent mixing, failed to} capture exactly the correct number of shock reflections for detonation re-initiation to occur.  Furthermore, the transverse detonations observed in the experiments have not been captured numerically.  In order to address turbulent mixing in DDT problems, through inclusion of molecular diffusion effects, some recent investigations have attempted direct numerical simulation (DNS) of the governing Navier-Stokes equations \cite{Gamezo2008,Ogawa2013,Houim2016,Poludnenko2017}. To capture the correct reaction rates and DDT event locations, however, problems are limited to the micro-scale and often include simplified chemistry considerations.  This is due to the required resolution to capture mixing on the molecular scale.  To address this limitation, LES is currently viewed as a reasonable compromise between accuracy of solution and resolvability of the problem. To date, an LES investigation using a \emph{flamelet} approach \cite{Johansen2013} has been applied to model the initial states of flame acceleration due to flame interaction with obstacles.  Unfortunately, the flamelet approach was not appropriate for capturing the correct reaction rate in the later stages of flame acceleration and DDT, as turbulent fluctuations were expected to increase beyond the flamelet combustion regime.  To address this, Gaathaug et al. \cite{Gaathaug2012} have applied a hybrid LES strategy to model DDT of hydrogen-air mixtures which treats combustion in both the extreme limits of \emph{flamelet} and perfectly mixed, \emph{well-stirred reactor}, regimes.  Unfortunately, detonation initiation events were observed to occur much sooner in the simulations compared to their corresponding experiments.  This is likely due to the fact that the hybrid method does not treat combustion rates when turbulent mixing and chemical reaction rates are comparable.  Finally, the \emph{flame-thickening} approach has also been attempted with some limited success \cite{Yu2015,Emami2015}.  In this regard, the flame-thickening approach was able to produce grid-independent and converged results for DDT timings and locations resulting from turbulent mixing. This approach, however, has yet to be validated against experiment.  Also, the flame-thickening approach relies on the compromise of the sensitivity of the gas to shock-compression, and hence the reaction rate, in order to capture the correct laminar flame speed.  This may, therefore, be problematic in predicting accurately the onset of DDT events, quantitatively, for a given fuel mixture.

More recently, experiments at the University of Ottawa \cite{Ahmed2016,Saif2016} have correlated detonation re-initiation events to a stability criterion, the $\chi$ parameter \cite{Radulescu2013}, which is the product of the mixture activation energy and the ratio of chemical induction to reaction time.  The transition length to initiate a self-sustained detonation was found to correlate very well with the mixtures’ sensitivity to temperature fluctuations.  Thus, it was found that DDT events were more likely to occur as the mixture irregularity increased.  Furthermore, it was also found that a necessary condition for DDT was the acceleration of the flame to the critical CJ \emph{deflagration} velocity \cite{Saif2016}.  Since irregular mixtures contain highly turbulent flow fields, it is therefore likely that turbulent mixing is a dominant mechanism that influences the DDT process.  To what extent, however, is  the topic of investigation below.

In the current work, detonation attenuation by a porous medium, as depicted in Figure \ref{fig.Radulescu2011_exp}, and the subsequent re-initiation was modelled, numerically, using the compressible LEM-LES (CLEM-LES) approach \cite{Maxwell2016}. This approach is a grid-within-a-grid approach, based on the linear eddy model for large eddy simulation (LEM-LES) \cite{Menon2011}.  The CLEM-LES was recently validated to experiments and applied to investigate the role of \emph{turbulent mixing} on unobstructed, irregular detonation propagation in a narrow channel filled with premixed methane-oxygen at low pressures \cite{Maxwell2016,Maxwell2016b}.  In this recent investigation, it was found that altering the turbulent mixing rates had a significant impact on the detonation hydrodynamic structure, cell size, and formation of un-burned pockets in the wake.  In the current study, the same approach was adopted:  To validate DDT events observed in numerical simulations with experimental observations, and to determine how such events and local reaction rates are influenced by changes in the turbulent fluctuations present.

\section{A Review of the Validating Experiments}

\subsection{Methodology}

\begin{figure}[ht]
\centering
\includegraphics[scale=0.75]{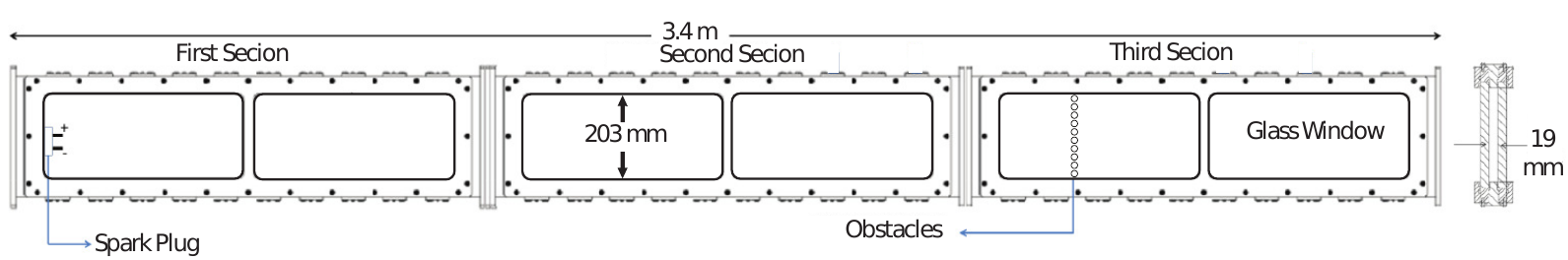}
\caption{Schematic of the experimental set-up \cite{Ahmed2016}.}
\label{fig.Ahmed2016_exp}
\end{figure}

For the experiments conducted in references \cite{Maley2015,Ahmed2016,Saif2016}, whose data serves to validate the numerical investigation conducted here, a shock tube technique was used, as illustrated in Fig.\ \ref{fig.Ahmed2016_exp}.  The shock tube was 3.4 m in total length and had a rectangular cross section whose height was 203 mm by 19 mm wide.  The narrowness of its cross section in one direction permitted the establishment of flow fields with high aspect ratios whose flow structure was essentially two-dimensional.  A large-scale Edgerton shadowgraph technique \cite{Settles2001} was implemented using a 2 m by 2 m retro-reflective screen, a hig-speed Phantom v1210 camera, and a Xenon arc continuous light source \cite{Dennis2014}.  Images were obtained from experiments at a frame rate of 42,049 fps using 1152$\times$256 pixels resolution, which corresponded to temporal and spatial resolutions of approximately 23.8 {\gmu}s and 1 mm, respectively.   A row of 5-10 cylindrical obstacles were placed within the test section near the entrance to an optical access window.  This permitted visualization of the downstream fast flame establishment, which corresponded to the final states of DDT.  The cylinders were sized such that the total blockage ratio of the shock tube was 75\%.  In all cases, the shock tube was evacuated to pressures below 80 Pa prior to filling with the desired test mixture.  Finally, to ignite the test mixture, a high voltage spark was obtained from a capacitor discharge yielding 1 kJ with a deposition time of 2 {\gmu}s.

\subsection{Observations of flame acceleration and DDT}

\begin{figure}[ht]
\centering
\includegraphics[scale=1.0]{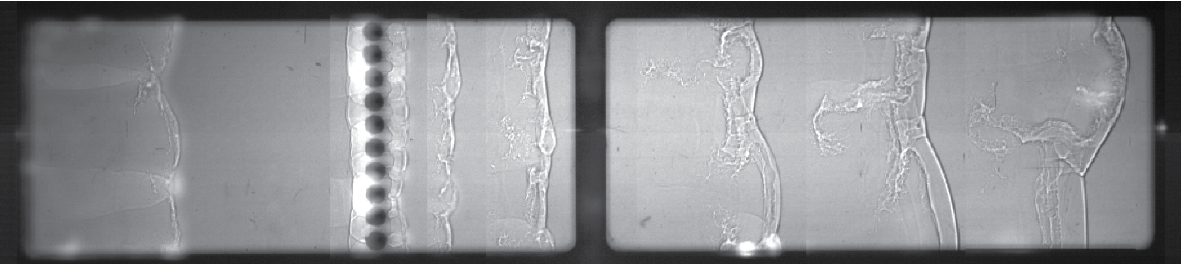}
\caption{Superimposed shadowgraph image frames showing the evolution of the fast flame in $\textrm{CH}_4+\textrm{2O}_2$ at $\hat{p}_o=8.2$ kPa with a 75\% blockage ratio \cite{Saif2016}.}
\label{fig.Saif2016_exp}
\end{figure}

A number of experiments were carried out for various fuel-oxygen mixtures at various pressures.  An example shadowgraph flow field evolution from one of these experiments is shown Fig.\ \ref{fig.Saif2016_exp} for a case where the fast-flame accelerated following the detonation interaction with the porous medium in stoichiometric methane-oxygen at 8.2 kPa.  In this figure, only selected images from Saif et al. \cite{Saif2016} are shown for a wave travelling from left to right.  Immediately following the interaction of the detonation wave with the porous medium, an irregular flow structure was observed.  At most locations on the wave front, the reaction zone was decoupled from the leading shock wave.  However, some local explosion events were observed in locations where Mach shock reflections have occurred.  These local explosion events were largely believed to arise due to the combination of intense shock compression near the triple point, and also forward jetting of hot combustion products along turbulent shear layers behind the leading Mach shock through the Kelvin-Helmholtz instability \cite{Maley2015}.  As the wave front evolved, the structure appeared to enter a lower mode of oscillation.  The cellular pattern appeared to grow in size.  As a result, there were fewer triple points, larger distances between shocks and reaction zones, and fewer local explosion events.  Also, despite the occurrence of these local explosion events, and similar hydrodynamic flow structure for self-sustaining detonation wave propagation \cite{Radulescu2007b,Maxwell2016b}, the overall wave velocity was found to be around 35\% lower than the theoretical CJ detonation velocity.  In some cases, DDT occurred and the wave structure resumed it's characteristic fine-scale cellular structure with near CJ-wave velocity magnitudes.

\begin{figure}[ht]
	\centering
	\includegraphics[scale=1.0]{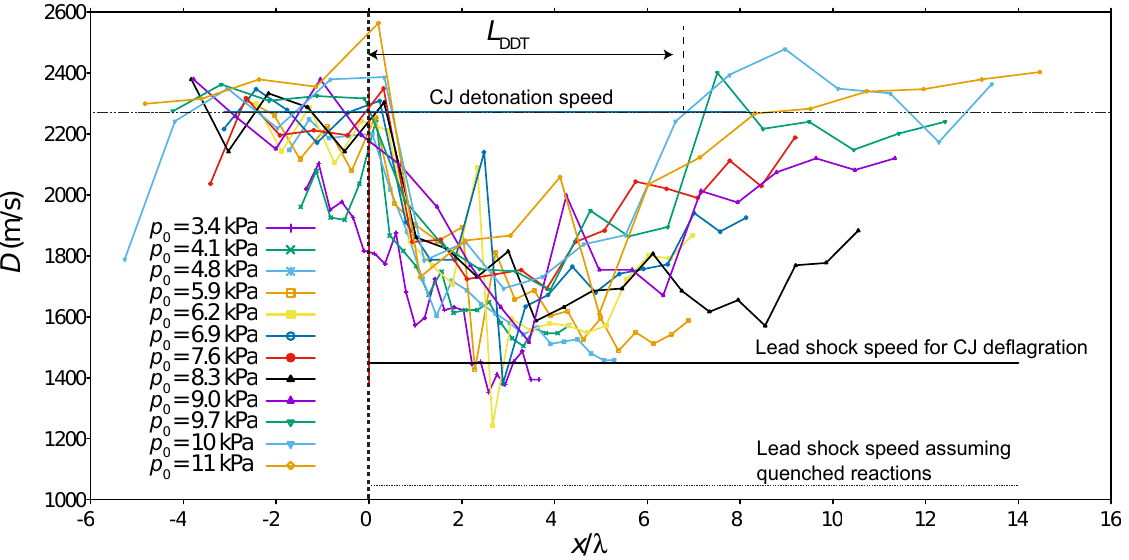}
	\caption{Lead shock speed variation with the distance travelled from the location of the column of cylinders for the
		experiments conducted in $\textrm{CH}_4+\textrm{2O}_2$ using the 75\% blockage ratio \cite{Saif2016}.}
	\label{fig.Ahmed2016_exp2}
\end{figure}

Velocity measurements of the average leading shock wave positions were obtained for fast-flame acceleration experiments in $\textrm{CH}_4+\textrm{2O}_2$, at various pressures, and are shown in Fig.\ \ref{fig.Ahmed2016_exp2}.  In this figure, the average leading shock wave position, for any given instance in time, was determined by locating the shock position at five equally spaced positions along the shock tube height and then ensemble averaged accordingly.  The wave velocities were then reported as a function of distance from the row of obstacles, which have been normalized by the corresponding mixture cell size, $\lambda$, which was determined from Shepherd's detonation database \cite{Kaneshige1997}.  In all cases where DDT was observed, for initial mixture pressures of $\hat{p}_o\ge9.7$ kPa, the distance to detonation transition was obtained at the locations, downstream from the obstacles, where the average wave velocities recovered the CJ detonation speed.  In general, the distance for DDT to occur was approximately $L_{\textrm{DDT}}\approx7\lambda$.  For the lowest pressures, $\hat{p}_o\le5.9$ kPa, the wave velocities were found to continually decrease with time and DDT was not observed. For the remainder of the experiments, however, upon an initial decrease in wave velocity, the wave was always observed to accelerate toward the CJ detonation speed.  From the full set of experiments across several different types of fuel, the CJ-deflagration speed was proposed as a required threshold velocity that must be reached or maintained in order for acceleration to detonation to occur \cite{Ahmed2016,Saif2016}.  This CJ-deflagration speed, shown in Fig.\ \ref{fig.Ahmed2016_exp2}, was determined from a closed form gas dynamic and thermodynamic model formulation of the one-dimensional choked deflagration structure \cite{Radulescu2015}.  In the remainder of the paper, advanced numerical modelling will attempt to confirm the role of this CJ-deflagration threshold on DDT, and to determine the roles of turbulent mixing and local burning rates on sustaining this threshold and, subsequently, the transition to detonation.

\section{Modelling Approach}

\subsection{The filtered LES equations}

For highly compressible and transient flows, high Mach numbers ($M_a$) and Reynolds numbers ($R_e$) are expected.  As a result, a wide range of length scales, governed by the compressible Navier-Stokes equations, must be resolved.  Since direct numerical simulation of the Navier-Stokes equations is extremely prohibitive, the approach adopted here has been to filter the governing equations through the LES methodology.  Rapid transients and fluid motions were thus captured on the large scales, while the small scale contributions were modelled through source terms. For a calorically perfect fluid system, The LES-filtered conservation equations for mass, momentum, and energy are given below in Eqs.\ \eqref{eqn.LESmass} through \eqref{eqn.LESenergy}, respectively.  A one-equation Localized Kinetic energy Model (LKM) was also used to describe the evolution of sub-grid velocity fluctuations in the form of sub-grid kinetic energy ${k}^{sgs}$, see Eq.\ \eqref{eqn.ke}.  Finally, the equations of state are given by Eq.\ \eqref{eqn.EOS}.  The equations below are given in non-dimensional form, where the various gas properties were normalized by the reference quiescent state. Favre-average (LES) filtering was achieved by letting $\tilde{f} = \overline{\rho f} / \bar{\rho}$, where $f$ represents one of the many state variables.  Here $\rho$, $p$, $e$, $T$, and $\boldsymbol{u}$ refer to density, pressure, specific sensible + kinetic energy, temperature, and velocity vector, respectively.
 
\footnotesize
\begin{gather}
   \frac{\partial \bar{\rho}}{\partial {t}} + \nabla \cdot (\bar{\rho} \tilde{{\boldsymbol{u}}}) = 0
   \label{eqn.LESmass} \\
   \frac{\partial \bar{\rho} \tilde{{\boldsymbol{u}}}}{\partial {t}} + \nabla \cdot (\bar{\rho} \tilde{\boldsymbol{u}} \tilde{{\boldsymbol{u}}}) + \nabla \bar{p} - \nabla \cdot {\bar{\rho}(\nu + \nu_{t})} \biggl( \nabla \tilde{\boldsymbol{u}} + (\nabla \tilde{\boldsymbol{u}})^T - \frac{2}{3}( \nabla \cdot \tilde{{\boldsymbol{u}}} ) \hat{I} \biggr) = 0
   \label{eqn.LESmomentum}
\end{gather}
\begin{gather}
   \frac{\partial \bar{\rho} \tilde{e}}{\partial {t}} + \nabla \cdot \biggl((\bar{\rho}\tilde{e}+\bar{p})\tilde{{\boldsymbol{u}}} - \tilde{{\boldsymbol{u}}} \cdot \bar{{\tau}}\biggr) - \biggl(\frac{\gamma}{\gamma - 1} \biggr) \nabla \cdot \biggl( \bar{\rho} (\frac{\nu}{P_r} + \frac{\nu_t}{P_{r,t}}) \nabla \tilde{T} \biggr) = - Q \overline{\dot{\omega}}
   \label{eqn.LESenergy} \\
   \frac{\partial \bar{\rho} {{k}^{sgs}}}{\partial {t}} + \nabla \cdot (\bar{\rho} \tilde{{\boldsymbol{u}}} {{k}^{sgs}}) - \nabla \cdot \biggl({\frac{\bar{\rho} \nu_{t}}{P_{r,t}} \nabla {{k}^{sgs}}} \biggr) = \bar{\rho}\nu_{t} \biggl( \nabla \tilde{\boldsymbol{u}} + (\nabla \tilde{\boldsymbol{u}})^T - \frac{2}{3}( \nabla \cdot \tilde{{\boldsymbol{u}}} ) \hat{I} \biggr) \cdot (\nabla \tilde{\boldsymbol{u}}) - \bar{\rho} \epsilon
\label{eqn.ke} \\
%
   \tilde{e} = \frac{{\bar{p}}/{\bar{\rho}}}{(\gamma - 1)} + \frac{1}{2} \tilde{\boldsymbol{u}} \tilde{\boldsymbol{u}} + \frac{1}{2} {{k}^{sgs}} \;\;\;\;\;\;\;\; \mbox{and} \;\;\;\;\;\;\;\;  \bar{\rho} \tilde{T} = \bar{p}
   \label{eqn.EOS}
\end{gather}
\normalsize
\noindent Non-dimensionalization of the various state variables has been achieved through
\begin{equation}
\rho = \frac{\hat{\rho}}{\hat{\rho_o}}, \;\;\; \boldsymbol{u} = \frac{\hat{\boldsymbol{u}}}{\hat{c_o}},  \;\;\; p = \frac{\hat{p}}{\hat{\rho_o} {\hat{c_o}}^2} = \frac{\hat{p}}{\gamma \hat{p_o}}, \;\;\; T = \frac{\hat{T}}{\gamma \hat{T_o}}, \;\;\; x = \frac{\hat{x}}{\hat{\Delta}_{1/2}}, \;\;\; t = \frac{\hat{t}}{\hat{\Delta}_{1/2} / \hat{c_o}}
\label{eqn.nonDim1}
\end{equation}
\noindent where the subscript `o' refers to the reference state, the hat superscript refers to a dimensional quantity, $c$ is the speed of sound, and $\hat{\Delta}_{1/2}$ is a reference length scale.  This reference length scale is taken as the theoretical half-reaction length associated with the steady Zeldovich, Von Neumann, and Doring (ZND) detonation wave solution \cite{Fickett1979} in the quiescent reference fluid. Other usual properties to note are the heat release, $Q$, the ratio of specific heats, $\gamma$, the kinematic viscosity, $\nu$, and the identity matrix, $\hat{I}$. The turbulent viscosity and dissipation were modelled according to
\begin{equation}
\nu_{t} = \frac{1}{\pi}\biggl(\frac{2}{3C_\kappa}\biggr)^{3/2}\sqrt{k^{sgs}}\bar{\Delta}
\label{eqn.turbulentViscosity}
\end{equation}%
and
\begin{equation}
\epsilon = {\pi\biggl(\frac{2{k}^{sgs}}{3C_\kappa}\biggr)^{3/2}}/{\bar{\Delta}}
\label{eqn.turbulentDissipation}
\end{equation}%
respectively. Here, $\bar{\Delta}$ was the minimum grid spacing, and $C_{\kappa}$ was the \emph{Kolmogorov} number, a model parameter which required calibration.  Finally, the chemical reaction term, $\overline{\dot{\omega}}$, required closure.

A second order exact Godunov compressible flow solver \cite{Falle1991} was applied to evolve the system of Eqs.\ \eqref{eqn.LESmass} through \eqref{eqn.ke} on Cartesian grids.  Adaptive mesh refinement (AMR) \cite{Falle1993} was also applied to increase computational efficiency by resolving only the regions where shocked unreacted gas was present.  See Maxwell \cite{Maxwell2016} for specific details.

\subsection{LEM subgrid combustion model}

In order to close the chemical reaction term, $\overline{\dot{\omega}}$, the CLEM sub-grid modelling strategy was applied \cite{Maxwell2015}.  Here, the micro-scale mixing and chemical reaction were handled entirely on the sub-grid, through a supplementary simulation of a 1D sample of the flow field within each fully refined LES cell.  The system of equations that were solved on the sub-grid was the conservation of enthalpy, Eq.\ \eqref{eqn.LEMenergy}, and the conservation of reactant mass, Eq.\ \eqref{eqn.LEMreactant}.  The source terms, $\dot{F_T}$ and $\dot{F_Y}$ accounted for the effect of turbulence on the sub-grid in the form of random ``stirring" events \cite{Menon2011} and $\dot{p}$ accounted for the energy changes associated with rapid changes in pressure, which are obtained entirely from the large-scale simulation, Eqs.\ \eqref{eqn.LESmass} to \eqref{eqn.ke}.  Next, $m$ was a one-dimensional mass weighted coordinate whose transformation to Cartesian spatial coordinates is given by Eq.\ \eqref{eqn.transform}.  Finally, a one-step Arrhenius combustion model was assumed through Eq.\ \eqref{eqn.reactRate}, which used a single reactant species, with mass fraction $Y$.  Also, $E_a$ and $A$ are the respective activation energy and pre-exponential factor model parameters required for the one-step combustion model.  Full details of the procedure, including the pressure coupling and LEM stirring, are found elsewhere \cite{Maxwell2016}.

\vspace*{-0.25in}
\begin{gather}
   \rho \frac{D T}{D t} - \biggl(\frac{\gamma - 1}{\gamma} \biggr)\dot{p} - \rho \frac{\partial}{\partial m}\biggl(\rho^2 \frac{\nu}{P_r} \frac{\partial T}{\partial m} \biggr) = - \biggl(\frac{\gamma - 1}{\gamma} \biggr) Q \dot{\omega} + \dot{F_T}
\label{eqn.LEMenergy} \\
   \rho \frac{D Y}{D t} - \rho \frac{\partial}{\partial m}\biggl(\rho^2 \frac{\nu}{L_e P_r} \frac{\partial Y}{\partial m} \biggr) = \dot{\omega} + \dot{F_Y}
\label{eqn.LEMreactant} \\
   m(x,t) = \int_{x_o}^x{\rho(x,t)dx}
\label{eqn.transform} \\
   \dot{\omega} = - {\rho} A Y e^{(-E_a / T)} 
\label{eqn.reactRate}
\end{gather}

\subsection{Numerical domain and model parameters}
\label{sec.param}

\begin{figure}[ht]
\centering
\includegraphics[scale=0.36]{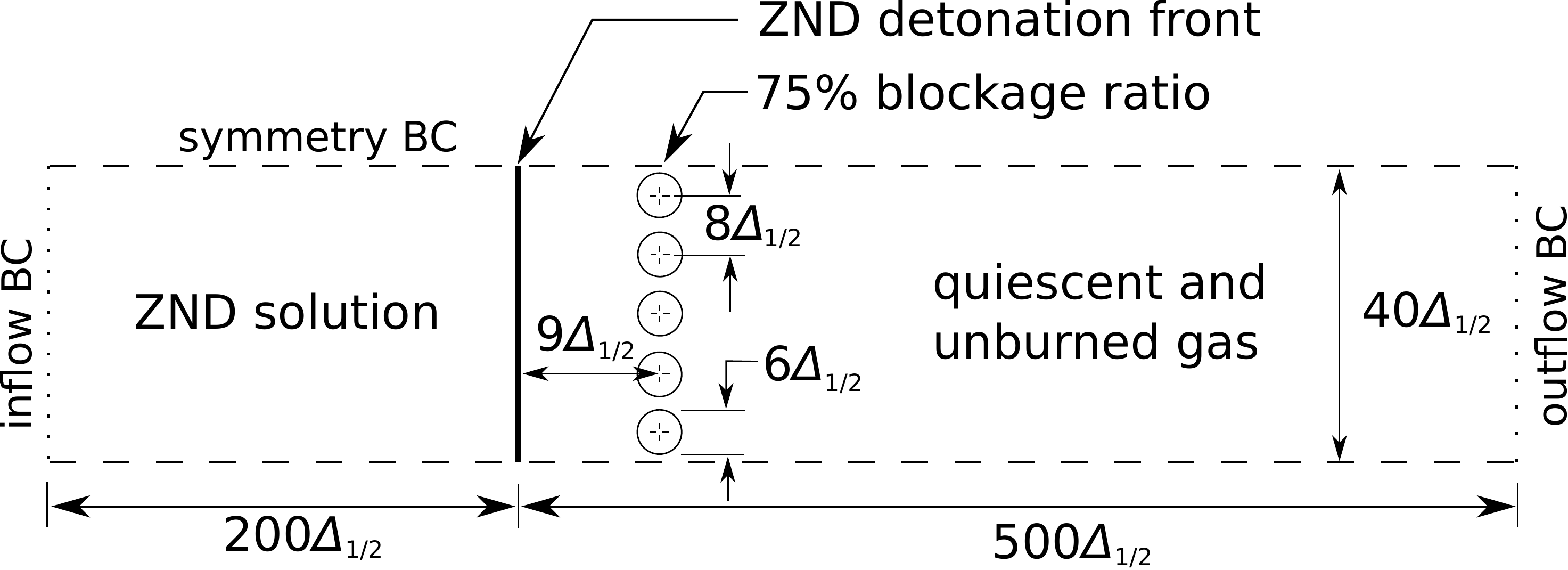}
\caption{Numerical domain.}
\label{fig2}
\end{figure}

A two-dimensional domain was considered where detonation waves were first attenuated using a bank of five cylinders with a blockage ratio of 75\%, as illustrated in Fig.\ \ref{fig2}.  The initial conditions and model parameters were consistent with physical experiments \cite{Maley2015,Ahmed2016,Saif2016}.  Owing to the predominantly two-dimensional large scale flow field evolution which arose from the high aspect ratio of the experimental apparatus, previously shown in Fig.\ \ref{fig.Ahmed2016_exp}, {three-dimensional simulations were not considered here.  For high aspect ratio flows, two-dimensional CLEM-LES investigations were previously found to reproduce well the statistical veolicity distribution behaviour and hydrodynamic structure of detonation propagation when compared to three-dimensional simulations, with much less computational expense} \cite{Maxwell2016b}. The steady ZND detonation wave solution was imposed 9 half-reaction lengths ($9\Delta_{1/2}$) upstream from the bank of cylinders.  The test section measured $500\Delta_{1/2}$ by $40\Delta_{1/2}$, which was comparable to the physical experiments \cite{Maley2015,Ahmed2016,Saif2016}.  Symmetric boundary conditions were imposed on the top and bottom walls, with inlet and outlet boundary conditions at the domain ends as shown.  The inlet boundary was sufficiently far from the test section ($200\Delta_{1/2}$) such that the results were unaffected by its influence.

The model parameters, $Q$ and $E_a$, were tuned to reproduce the correct post-shock ignition delay times for premixed methane-oxygen.  Also, the pre-exponential factor, $A$, and diffusion coefficients were {chosen} such that the one-step model reproduced the correct half reaction length, $\hat{\Delta}_{1/2}$, and also the {correct laminar premixed flame speed at post-shock conditions, prior to auto-ignition,} for a shock travelling at 70\% the theoretical CJ speed for the given quiescent mixture.  These parameters (ignition delay time, $\hat{\Delta}_{1/2}$, and flame speed) were determined from full chemistry considerations using Cantera \cite{Goodwin2013} and the GRI-3.0 kinetic mechanism \cite{Smith2013}.  See Maxwell et al. \cite{Maxwell2016b} for details.  The full set of dimensional and non-dimensional model parameters are shown in table \ref{tab:params_2dmethane}. The LES-scale resolution used was $\bar{\Delta}=\Delta_{1/2}/32$, with an additional 16 sub-grid elements within each LES cell, for an effective resolution of $\bar{\Delta}_{eff}=\Delta_{1/2}/512$.  This resolution was previously shown to resolve the post-shock laminar flame speeds and detonation structure for this particular methane-oxygen mixture \cite{Maxwell2016b}.  A grid convergence study was therefore not repeated here.  Finally, $C_\kappa$ was varied in order to change the amount of turbulent velocity fluctuations generated by the wave dynamics, and to consequently research their affect on DDT.

\begin{table*}
  \centering
  \caption{Dimensional and non-dimensional fluid properties and model parameters for methane combustion initially at $\hat{T}_o=300$K and $\hat{p}_o=11$kPa.}
  {\small
  \resizebox{\columnwidth}{!}{
  \begin{tabular}{llllllllll}
    \hline
    \multicolumn{6}{l}{\textbf{Dimensional properties}} \\
    \hline
    $\hat{\rho_o}$ & 0.12 kg $\textrm{m}^{-3}$    & & $\hat{c_o}$        & 356.36 m $\textrm{s}^{-1}$ & & $\hat{E_a}/\hat{R}$   & 18746.2 K  \\
    $\hat{D}_{CJ}$ & 2292.53 m $\textrm{s}^{-1}$                  & & $\hat{D}_{70\%CJ}$ & 1604.77 m $\textrm{s}^{-1}$ & & $\hat{S}_{L,70\%CJ}$  & 16.46 m $\textrm{s}^{-1}$\\
    $\hat{T}_{CJ}$ & 3337.39 K                    & & $\hat{p}_{CJ}$     & 296.8 kPa    & & $\hat{\rho}_{CJ}$     & 0.22 kg $\textrm{m}^{-3}$  \\
    $\hat{\nu}$    & 6.1x$10^{-5}$ $\textrm{m}^2$ $\textrm{s}^{-1}$ & & $\hat{k}/{(\hat{\rho}\hat{c_p})}$ & 1.6x$10^{-4}$ $\textrm{m}^2 \textrm{s}^{-1}$ & & $\hat{D}$ & 6.5x$10^{-5}$ $\textrm{m}^2$ $\textrm{s}^{-1}$  \\
    $\hat{Q}$             & 6754.7 kJ $\textrm{kg}^{-1}$     & & $\hat{\Delta}_{1/2}$   & 2.48 mm  & & $\hat{\lambda}$ & 41-47 mm \cite{Kaneshige1997}\\

    \hline
    \multicolumn{6}{l}{\textbf{Non-dimensional model parameters}} \\
    \hline
    $\nu$ & 4.14x$10^{-3}$ & & $D_{CJ}$ & 6.43 & & $D_{70\%CJ}$ & 4.50 \\
    $Le$ & 1.32 & & $Pr$ & 0.709 & & $Sc$ & 0.933 \\
    $Pr_t$ & 1.0 & & $Sc_t$ & 1.0 & & $\gamma$ & 1.17 \\
    $E_a$ & 46.0 &  & $Q$ & 53.2 & & $A$ & $7.23$x$10^3$  \\
    
  \end{tabular}}
  }
  \label{tab:params_2dmethane}
\end{table*}

\section{Results}

\subsection{Flow field evolutions}

\begin{figure}[ht!]
\raggedright
a) $C_\kappa=2.0$:
\includegraphics[scale=0.3]{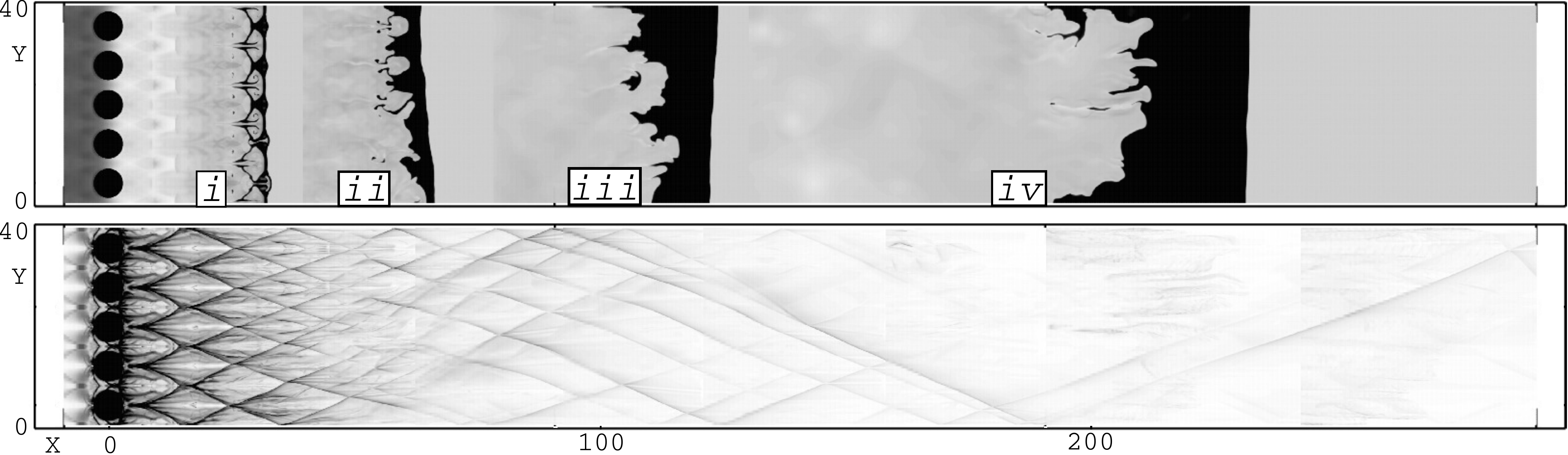}\\
b) $C_\kappa=2.6$:
\includegraphics[scale=0.3]{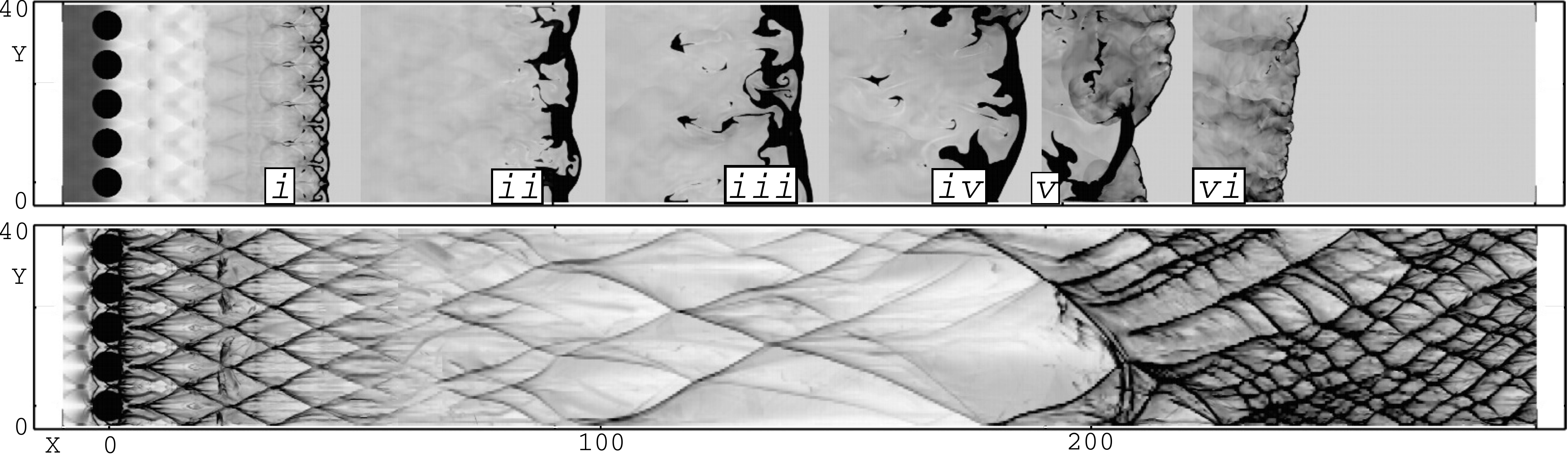}\\
c) $C_\kappa=3.0$:
\includegraphics[scale=0.3]{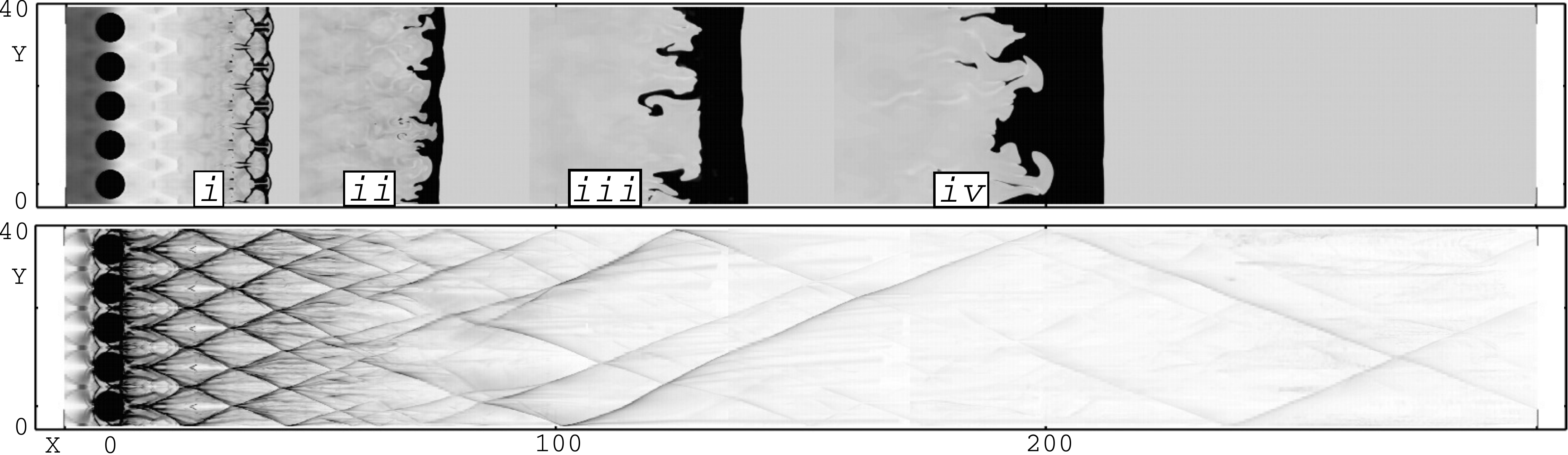}
\caption{Density evolution (top) and corresponding numerical soot foil (bottom) obtained for three $C_\kappa$ values ($C_\kappa=2.0$, 2.6, and 3.0).}
\label{fig.evolution}
\end{figure}

In this study, simulations were conducted for $C_\kappa$ values ranging from 2.0 to 3.0.  Flow field density evolutions are shown in Fig.\ \ref{fig.evolution} for three cases ($C_\kappa=2.0$, 2.6, and 3.0) where distinctly different flame acceleration behaviour was observed.  Also shown in Fig.\ \ref{fig.evolution} are the corresponding numerical soot foils, which were obtained by integrating the local vorticity at each spatial location, $\Omega(x,y)$, throughout the duration of the simulation through
\begin{equation}
\Omega(x,y) = \int_{t=0}^{t}\biggl(\boldsymbol{\nabla}\times\bar{\boldsymbol{u}}(x,y,t)\biggr)\textrm{d}t.
\label{eqn.sootFoil}
\end{equation}
In all three cases, detonation propagation is effectively quenched following the initial detonation interaction with the bank of cylinders.  This was observed by the decoupling of the flame from the leading shock wave, which lead to a thickening of the reaction zone, as seen in frame (\textit{i}) of {all three simulations in} Fig.\ \ref{fig.evolution}.  Furthermore, turbulent motions originating from the cylinders gave rise to the pattern observed in these frames.  For {$C_\kappa=2.0$}, subsequent frames in the density evolution (\textit{ii}-\textit{iv}) reveal an ever increasing thickening of the reaction zone as the wave travelled downstream from the obstacles.  In this case, detonation did not occur.  For {$C_\kappa=2.6$}, following the cellular observations in frame (\textit{i}), the turbulent instabilities intensified and give rise to a larger cellular structure as observed in frames (\textit{ii}) and (\textit{iii}).  This can be seen by the progressive increase of cell size with distance downstream from the obstacles in the corresponding numerical soot foil.  This tendency of the deflagration cell structure to enter larger modes was also observed experimentally \cite{Maley2015,Saif2016}, as shown previously in Fig.\ \ref{fig.Saif2016_exp}.  Also DNS investigations of O'Brien et al. \cite{Obrien2017} and Poludnenko \cite{Poludnenko2017} {have shown the same behaviour of turbulent flame acceleration.  They attributed this behaviour to the backscatter of kinetic energy, associated with fine-scale turbulent motions, to larger scales through flame generated baroclinic instabilities associated with the interactions of large pressure gradients with the expanding flame fronts.} By frame (\textit{iv}), the deflagration cell structure spanned the height of the domain.  At this point detonation occurred through collision of the shock triple point with the upper wall, around $x\approx170\Delta_{1/2}$, which gave rise to a close coupling between the reaction zone and leading shock, and also a much smaller and more prominent cellular pattern on the numerical soot foil.  In frame (\textit{v}), detonation initiated at a separate location along the bottom wall, around $x\approx180\Delta_{1/2}$.  Finally, by frame (\textit{vi}), the entire wave front existed as a detonation.  For {$C_\kappa=3.0$}, the same increase in cell size of the deflagration was observed downstream of the obstacles.  Despite this, the reaction zone continually increased in size as the wave evolved, and detonation did not occur.  Finally, upon comparing the numerical soot foils of all three simulations, increasing $C_\kappa$ had the effect of generating larger cells by the time the wave reached $x\approx80-100\Delta_{1/2}$.

\subsection{Wave velocity measurements}

\begin{figure}[ht!]
\centering
\includegraphics[scale=1.0]{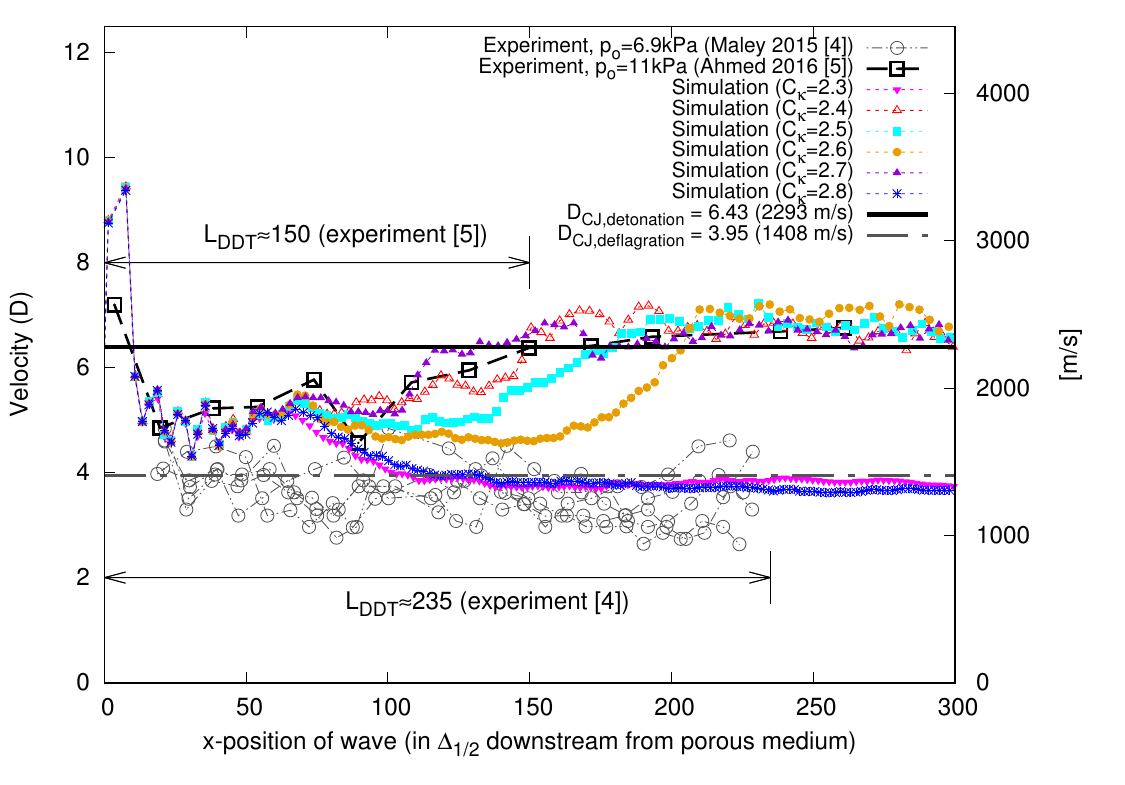}
\caption{Wave velocity, as a function of distance, for a range of $C_\kappa$.  Also shown are the correponding experimental results \cite{Maley2015,Ahmed2016}, obtained for $\textrm{CH}_4+\textrm{2O}_2$ at $\hat{p}_o=6.9$ kPa and $\hat{p}_o=11.0$ kPa, respectively.}
\label{fig.detspeeds}
\end{figure}

The re-initiation of detonation in each simulation was confirmed by measuring the average wave velocities, as a function of distance downstream from the obstacle centres, as shown in Fig.\ \ref{fig.detspeeds}.  In all cases, for $2.0\le C_\kappa \le 3.0$, the average wave velocity was determined by first locating the average wave locations at each time step.  These average wave locations were found by determining the $x$-locations where $\rho\ge1.1\rho_o$ at five equally spaced positions along the channel height and ensemble averaged accordingly.  For $2.4\le C_\kappa\le2.7$, the initially observed average wave speeds are found to approach but remain above the CJ-deflagration speed of $D=3.95$ (1408 m/s), as indicated in the figure.  This CJ-deflagration speed was determined using the method outlined in Radulescu et al. \cite{Radulescu2015}. Eventually, detonation occurred in these cases, and the waves travelled at a much faster velocity; the CJ detonation velocity of $D_{CJ}=6.43$ (2293 m/s).  The results from the corresponding physical experiments \cite{Maley2015,Ahmed2016} are also shown in Fig.\ \ref{fig.detspeeds}.  In Maley \cite{Maley2015}, the quenched wave was observed to travel near the CJ-deflagration speed, and showed signs of acceleration to the CJ-deflagration value when $x>200\Delta_{1/2}$.  In fact, detonation was observed at $x\approx235\Delta_{1/2}$, near the end of the experimental test section.  Unfortunately, velocity recordings beyond this point are not available.  However, Ahmed's experiment \cite{Ahmed2016} showed the same trend as the numerical simulations, where the initial wave velocities corresponded to the CJ-deflagration speed and eventually accelerated to the CJ-detonation value by $x\approx150\Delta_{1/2}$.  Also shown in Fig.\ \ref{fig.detspeeds} are the results from simulations where $C_\kappa=2.3$ and 2.8.  In these cases, the wave speeds were found to drop below the CJ-deflagration value, and detonation did not re-initiate for the duration of the simulations.  Table \ref{tab:LDDT} summarizes the detonation distances found for all simulations within the range $2.4\le C_\kappa\le2.7$.  In all of these simulations, $132\le L_\textrm{DDT}\le206$ ($7\lambda\le L_\textrm{DDT}\le11\lambda$), with $C_\kappa=2.4$ matching well the detonation length found in Ahmed's experiment \cite{Ahmed2016}.

\begin{table}[ht!]
  \centering
	\caption{DDT distances obtained from the numerical simulations.}
	{\small
		\begin{tabular}{|l|c|c|c|c|c|c|c|c|c|c|c|}
			\hline
			$C_\kappa$  & $\le$2.3 & 2.4 & 2.5 & 2.6 & 2.7 & $\ge$2.8 \\
			\hline
			$L_{\textrm{DDT}}$  & - & 150$\pm$3 & 182$\pm$3 & 206$\pm$3 & 132$\pm$3  & -  \\
			  &  & ($\sim8\lambda$) & ($\sim10\lambda$) & ($\sim11\lambda$) & ($\sim7\lambda$)  &   \\
			
			\hline
		\end{tabular}}
		\label{tab:LDDT}
	\end{table}

\section{Discussion}

\subsection{Effect of turbulent mixing rates on DDT}
\label{sec.disco1}

\begin{figure}[ht!]
	\centering
	\includegraphics[scale=1.0]{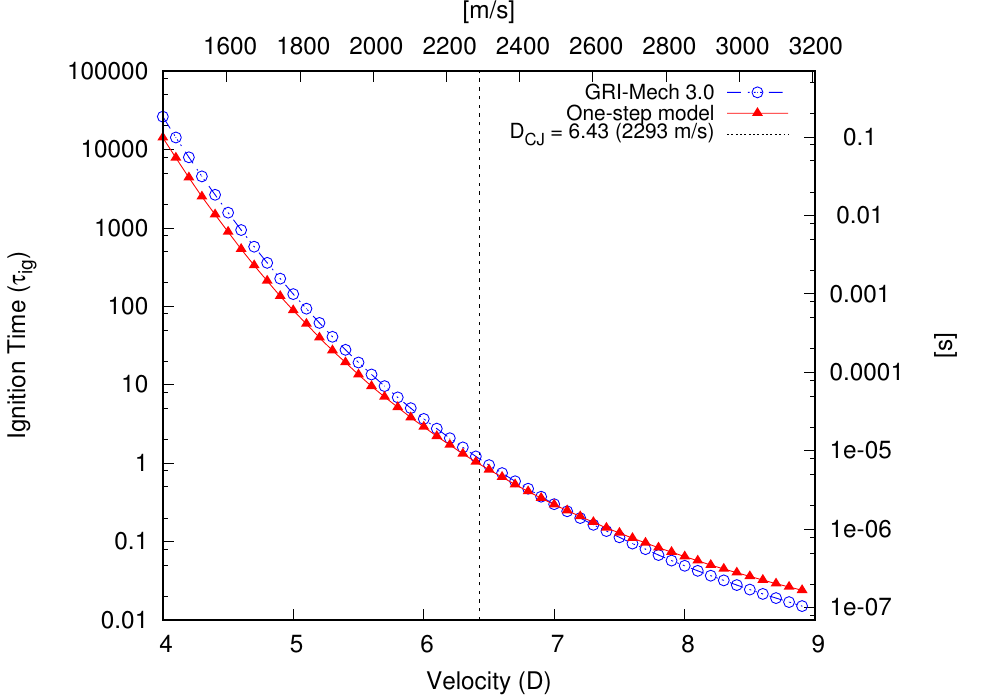}
	\caption{Ignition delay times for various shock speeds, computed using the one-step reaction mechanism, Eq.\ \ref{eqn.reactRate}, and also the GRI-3.0 mechanism \cite{Smith2013}.}
	\label{fig.ignitionTimes}
\end{figure}

In the current study, it is clear that sufficient turbulent mixing rates, through the appropriate tuning of $C_\kappa$, are required in order for DDT to occur.  When $C_\kappa\le2.3$, the wave speed was unable to maintain a velocity above the CJ-deflagration value, and DDT did not occur.  As $C_\kappa$ was increased to a value within the range $2.4\le C_\kappa\le2.7$, reaction rates at the surfaces of un-burned fuel in the wake of the wave front were sufficient to maintain the average wave velocity above the CJ-deflagration speed.  DDT then occurred in regions where transverse shock collisions occurred.  During these collision events, as will later be demonstrated in \cref{sec.disco4}, sufficient energy deposition due to shock compression coupled with intense turbulent burning allowed for a localized detonation to form, which eventually consumed the entire wave front.  As $C_\kappa$ was further increased beyond $C_\kappa\ge2.8$, DDT was once again mitigated.  It is noted that for all cases, as wave velocities drop below the CJ-detonation limit of $D_{CJ}=6.43$, the ignition delays associated with compression from the leading shock wave cannot sustain the flame front velocity.  For example, when the leading shock strength reduces to $D=5.0$ (1782 m/s), a constant volume ignition calculation reveals that the ignition delay for a particle passing through the wave front is on the order of $\tau_{ig}\approx100$ (0.001 s).  This can be seen in Fig.\ \ref{fig.ignitionTimes}, where constant volume ignition calculations have been carried out for shock strengths in the range $4\le D\le 9$ using both the one-step model and the GRI-3.0 kinetic mechanism \cite{Smith2013}.  Since the shocked particle velocity relative to the shock wave itself is nearly sonic, the distance to auto-ignition can be estimated from $\Delta_i\approx c\tau_{ig}$.  Thus, for this particular shock strength, $\Delta_i\approx100$ (250 mm).  Since the observed distances between the shock and flame are always much shorter, diffusive turbulent burning must be the dominant mechanism through which the flame accelerates prior to DDT.

\subsection{Local surface reactions and flame speeds}

\begin{figure}[ht!]
	\centering
	\includegraphics[trim=0cm 0cm 2.6cm 0cm, clip=true, scale=0.8]{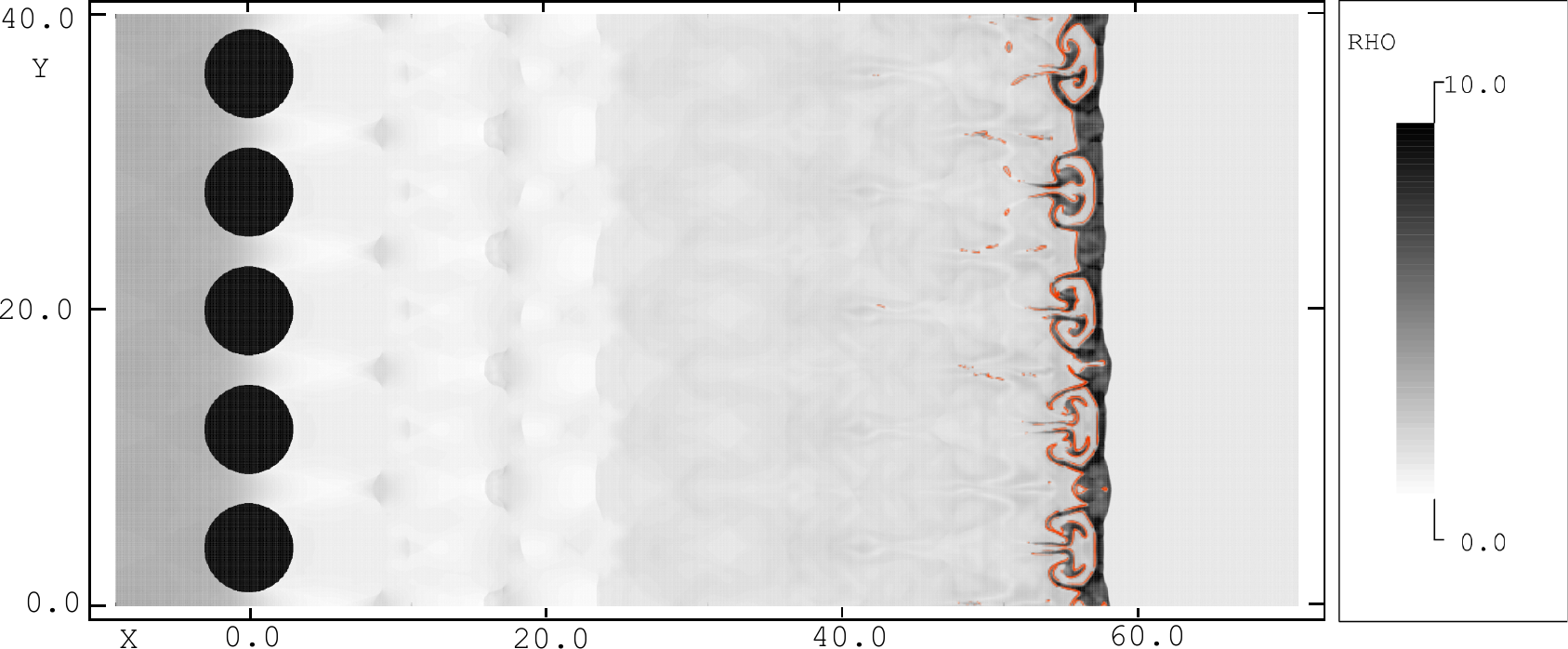}
	\caption{Instantaneous density field for $C_\kappa=2.4$ with superimposed reaction rate, $\dot{\omega}$, in red.}
	\label{fig.frame25}
\end{figure}

\begin{figure}[ht!]
	\centering
	\includegraphics[scale=1.0]{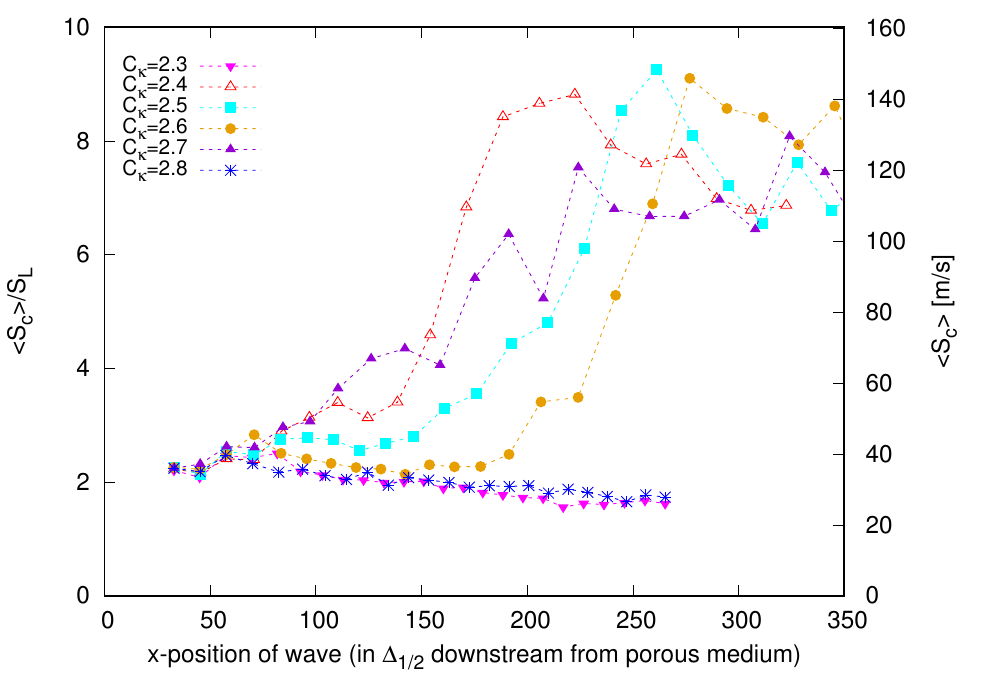}
	\caption{Average turbulent flame consumption speeds as a function of wave position downstream from the obstacles.}
	\label{fig.stvsx}
\end{figure}

In order to gain insight on how local surface reactions contribute to flame acceleration, it is useful to extract the reaction rate information which is readily available from simulation.  Figure \ref{fig.frame25} shows the instantaneous rate of reaction, $\overline{\dot{\omega}}$, which has been obtained from the $C_\kappa=2.4$ simulation and superimposed onto a corresponding density evolution plot.  From this figure, it is clear that the chemical reactions predominantly occur on the surfaces of burned-unburned gas interfaces, and not uniformly throughout shocked unbunred gas.  Since shock-compression cannot describe the locations of the surface reactions observed throughout the flow field, turbulent mixing is the dominant mechanism through which the unburned gases are consumed.  To further compliment this analysis, local flame consumption speeds have been obtained for all simulations where surface reactions have been found.  More specifically, these local flame consumption speeds have been determined on iso-contour locations where $Y=0.1$.  Accordingly, the local instantaneous flame consumption speeds have been evaluated from     
\begin{equation}
S_c=-\frac{1}{\rho_u Y_F}\int_{-\infty}^{\infty}\dot{\omega}\textrm{d}\boldsymbol{n}
\label{eqn.localFlameSpeed}
\end{equation}
where $\rho_u$ and $Y_F$ are the density and reactant mass fraction of the unburned fuel evaluated upstream from the flame surface, and $\boldsymbol{n}$ is the direction normal to the flame surface.  The instantaneous local values for $S_c$ have then been ensemble averaged at selected instances in time to give the evolution of the average local flame speed as the wave evolved.  The average flame consumption speeds, $\langle S_c\rangle$, are thus presented as functions of the wave position, $x$, downstream from the obstacles in Fig.\ \ref{fig.stvsx} for each simulation.  In all cases, the initial flame speeds of the quenched detonations, from $25<x<75$, were approximately two to three times the laminar flame speed ($S_L$).  For the cases where $2.4\le C_\kappa\le2.7$, these local flame speeds accelerated to values of $\langle S_c\rangle=7-9S_L$ by the time DDT occurred.  In fact, the increased magnitude of the local flame speed corresponded to the wave velocity trends shown previously in Fig.\ \ref{fig.detspeeds}.  At the onset of detonation, the increased averaged flame speed magnitudes, of 7 to 9 $S_L$ ({110-140 m/s}), were consistent with the local burning rates found in simulations and experiments of unobstructed detonation propagation in methane-oxygen \cite{Maxwell2016b}.  For $C_\kappa\le2.3$ and $C_\kappa\ge2.8$, the average local flame consumption speeds approached $\langle S_c\rangle\rightarrow2S_L$ as the waves evolved downstream from the obstacles.

\begin{figure}[ht!]
\centering
\includegraphics[scale=1.0]{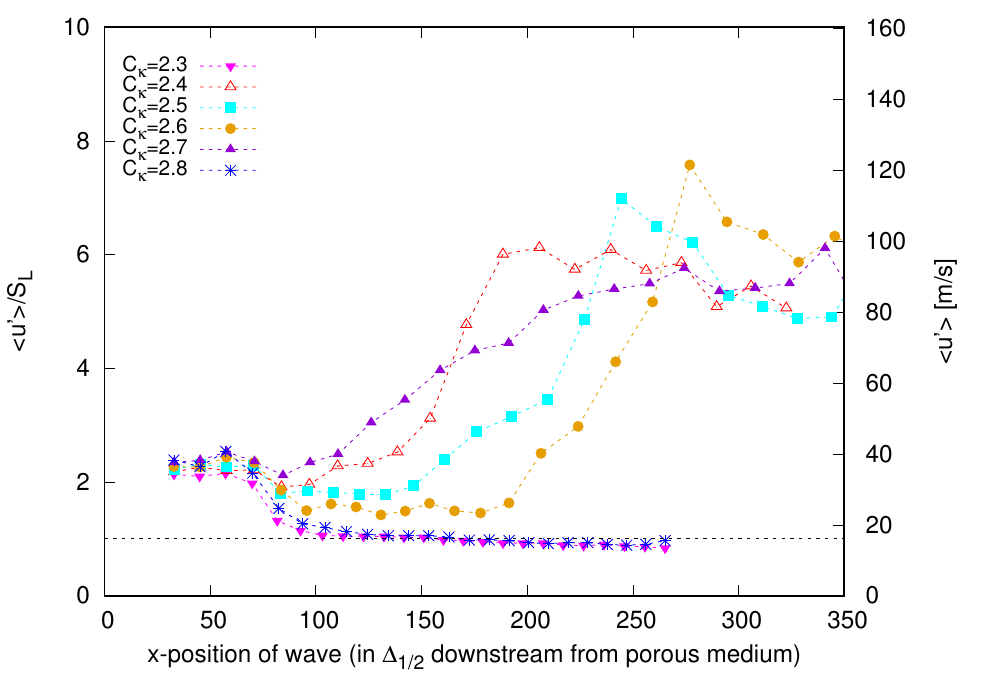}
\caption{Average turbulence intensity as a function of wave position downstream from the obstacles.}
\label{fig.upvsx}
\end{figure}

In order to understand the influence of turbulent mixing on the acceleration of local flame speeds, leading up to DDT, the averaged velocity fluctuation magnitudes ($\langle u'\rangle$) were also obtained from the flame surface iso-contours.  These are presented as functions of the wave position, $x$, downstream from the obstacles in Fig.\ \ref{fig.upvsx} for each simulation.  Here, the local turbulence intensities were determined from
\begin{equation}
u'=\sqrt{\frac{2}{3}k^{sgs}}.
\label{eqn.turbIntense}
\end{equation}%
In Fig.\ \ref{fig.upvsx}, the velocity fluctuations followed the same trends observed for the flame consumption speeds presented in Fig.\ \ref{fig.detspeeds}.  At first, $\langle u'\rangle\approx2S_L$ for all cases.  Around $x\approx75$ downstream from the obstacles, a reduction in $\langle u'\rangle$ was observed, which coincided with the formation of larger cells (or modes) seen previously in the numerical soot foils of Fig.\ \ref{fig.evolution}.  Despite this reduction in $\langle u'\rangle$, for $2.4\le C_\kappa\le2.7$, the local flame speeds in Fig.\ \ref{fig.stvsx} remained above $2S_L$.  Eventually, $\langle u'\rangle\rightarrow5$-$7S_L$ once detonation occurred.  Interestingly, for $C_\kappa\le2.3$ and $C_\kappa\ge2.8$, the averaged turbulence intensity dropped below $\langle u'\rangle=S_L$.  It is possible that velocity fluctuations above this threshold are required in order for flame acceleration to occur. It is, after all, well established that turbulent flame speeds are influenced by turbulent velocity fluctuations \cite{AbdelGayed1984}.  Also, it is interesting to note that increases in $C_\kappa$ do not necessarily lead to an increase in the turbulence intensity, $u'$, as previously demonstrated for unobstructed detonation propagation \cite{Maxwell2016b}.

\subsection{Global combustion regimes}

\begin{figure}[ht!]
\centering
\includegraphics[scale=1.0]{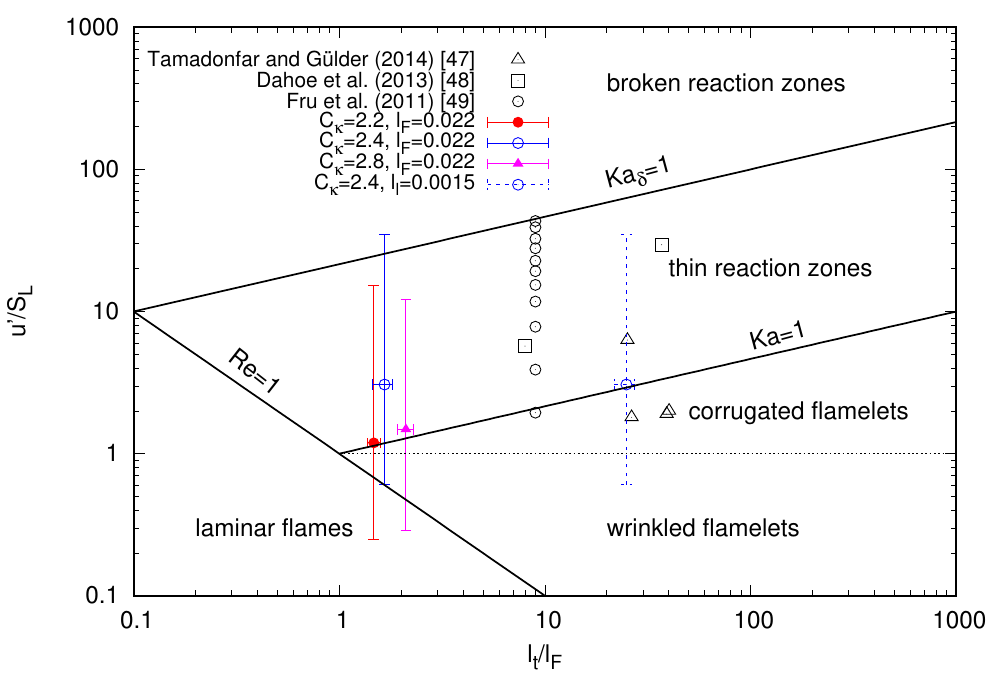}
\caption{Combustion regime diagram for selected simulations in methane-oxygen, and compared to experiments \cite{Tamadonfar2014,Dahoe2013} and DNS \cite{Fru2011} of turbulent methane-air flames.}
\label{fig.regime}
\end{figure}

In Fig.\ \ref{fig.regime}, the combustion regime diagram \cite{Peters1999} has been constructed for three of the simulations ($C_\kappa=2.2$, 2.4, and 2.8).  In this diagram, the average turbulence intensities relative to laminar flame speed ($u'/S_L$) and corresponding integral turbulent length scale relative to the flame thickness ($l_t/l_F$) are shown for all three simulations.  In addition to average values, the error bars show the extents of ($u'/S_L$) and ($l_t/l_F$) obtained in each simulation.  More specifically, the data shown has been obtained at locations where chemical reactions occur, i.e. where $\overline{\dot{\omega}}\ge0.01$, for the duration of each simulation.  Here, the turbulent length scale was defined as
\begin{equation}
l_t={u'}^3/\epsilon
\label{eqn.turbScale}
\end{equation}
and the flame length scale, associated with the inner structure of the laminar flame, was taken as
\begin{equation}
l_F=\frac{\Delta T}{\textrm{max}(\lvert{\nabla T}\rvert)}.
\label{eqn.FlameScale}
\end{equation}
For consistency with the model parameters previously determined in \cref{sec.param}, the laminar flame speed and thickness have been determined at the 70\% post-shock CJ condition.  For the one-step model applied in this investigation, $S_L=0.0462$ (16.46 m/s) and $l_F=0.022$ (55 \gmu m).  As previously observed in Fig.\ \ref{fig.upvsx}, the cases of $C_\kappa=2.2$ and $C_\kappa=2.8$ experienced much lower magnitudes of turbulence intensities ($u'$) compared to the case shown where DDT occurred ($C_\kappa=2.4$).  In fact, the range of $u'$ recorded for $C_\kappa=2.4$ extends all the way from the wrinkled flamelet regime to the broken reaction zones regime, while the other two cases shown extend only from the laminar flames regime to the thin reaction zones regime.  In all three cases, the average values lie within the thin reaction zones regime, which has significant implications on the use of available models for simulating DDT in methane-oxygen.  In general, increasing $C_\kappa$ was not found to increase the magnitude or range of $u'$, as previously observed in unobstructed detonation propagation \cite{Maxwell2016b}.  It was, however, found to increase the integral turbulent length scale, $l_t$.  Thus, increasing $C_\kappa$ generated larger eddies, and at some critical value, turbulent production terms dominated over local dissipation rates, which allowed for increased turbulence intensity, $u'$.  Also shown in Fig.\ \ref{fig.regime}, for comparison, are measurements of turbulent flames, in stoichiometric methane-air, obtained through experiments \cite{Tamadonfar2014,Dahoe2013} and direct numerical simulation with detailed chemistry \cite{Fru2011}.  While these investigations did not consider detonation, they did consider highly turbulent flames in the range of ($u'/S_L$) investigated here.  Although the ratio of integral turbulent length scales relative to the flame thickness ($l_t/l_F$) observed in these studies \cite{Tamadonfar2014,Dahoe2013,Fru2011} were found to be an order of magnitude larger than the current investigation, the combustion regimes were also found to lie predominantly within the thin reaction zones regime.  In fact, it is expected that realistic chemistry considerations would have a smaller $l_F$, owing to the fact that the laminar flame structures obtained from the one-step model are known to be much larger, with less steep gradients, than those obtained through detailed chemistry \cite{Maxwell2016b}.  For example, laminar flame simulations in methane-oxygen, at the 70\% post-shock CJ condition, using Cantera \cite{Goodwin2013} and the GRI-3.0 kinetic mechanism \cite{Smith2013}, are found to yield a flame width of only $l_F=0.0015$ (3.7 \gmu m).  Thus, upon rescaling the data obtained for $C_\kappa=2.4$, the range of ($l_t/l_F$) falls within the same order of magnitudes as the referenced literature \cite{Tamadonfar2014,Dahoe2013,Fru2011}, as shown in Fig.\ \ref{fig.regime}.  This re-scaling also lies within the corrugated flamelets and thin reaction zones regimes.  Finally, it is worth noting that LES which accounts for a more realistic flame structure would be recommended in future work to determine the exact extents of the combustion regimes expected for DDT of methane-oxygen.  This could be achieved either through inclusion of detailed chemistry, a 3 step model \cite{Short1997,Sharpe2006}, or a modified one-step model with induction zone kinetics \cite{Fickett1971} to better control the reaction zone thickness.  It is noted, however, that any of these three chemistry models would add considerable expense to the computation owing to the presence of stiff chemistry.

\subsection{The onset of detonation initiation}
\label{sec.disco4}

\begin{figure}[ht!]
	\centering
	\includegraphics[scale=0.55]{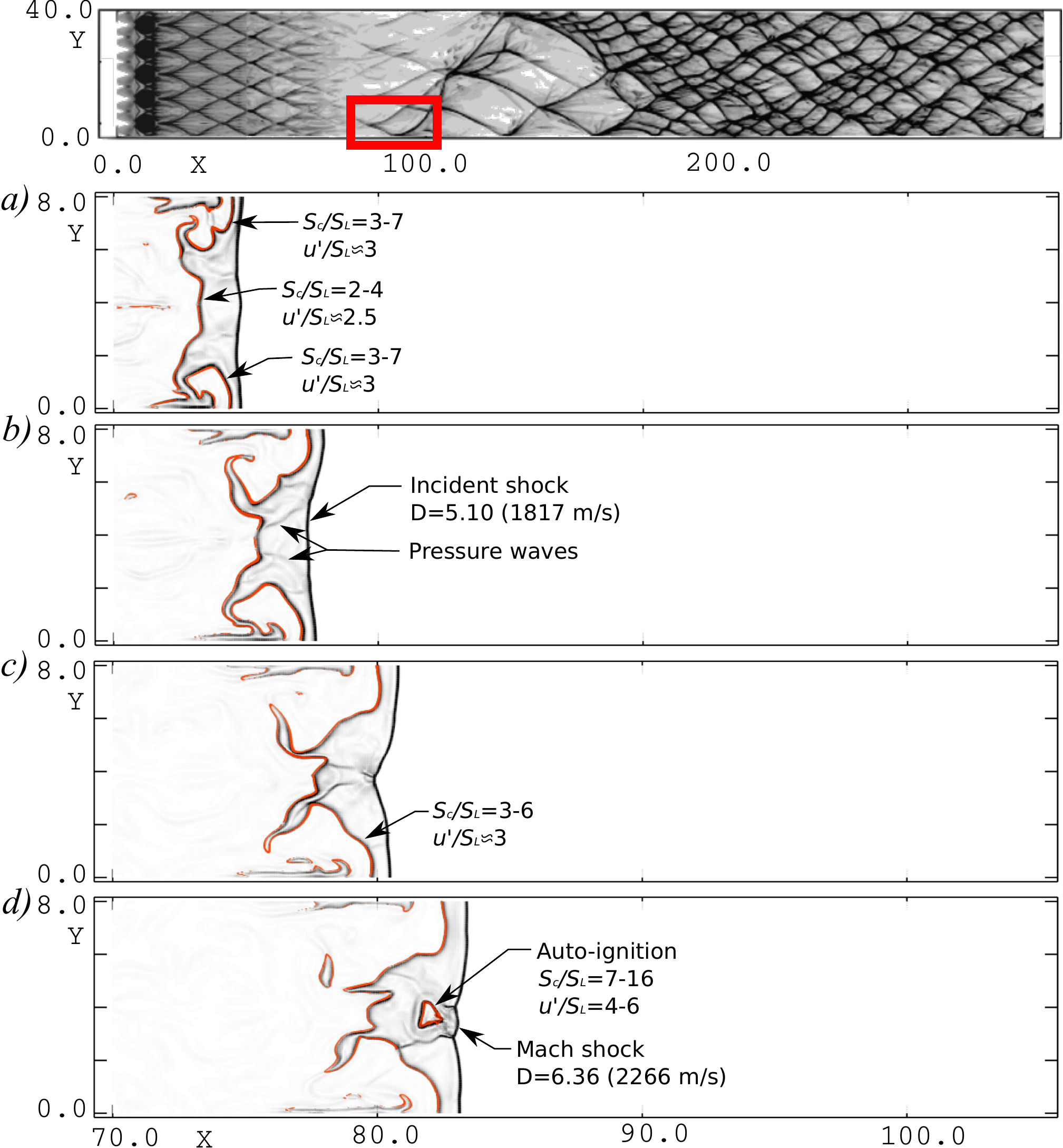}
	\caption{Onset of detonation initiation for $C\kappa=2.4$.  Frames (\textit{a}) through (\textit{d}) show the sequence of numerically obtained Schlieren images at $\Delta t=0.5$ (3.5 {\gmu}s) intervals for the region indicated on the numerical soot foil (top).  The local reaction rate, $\dot{\omega}$, has been superimposed onto the frames in red. Frames (\textit{e}) through (\textit{i}) are continued in Fig.\ \ref{fig.detinit2}.}
	\label{fig.detinit}
\end{figure}

\begin{figure}[ht!]
	\centering
	\includegraphics[scale=0.55]{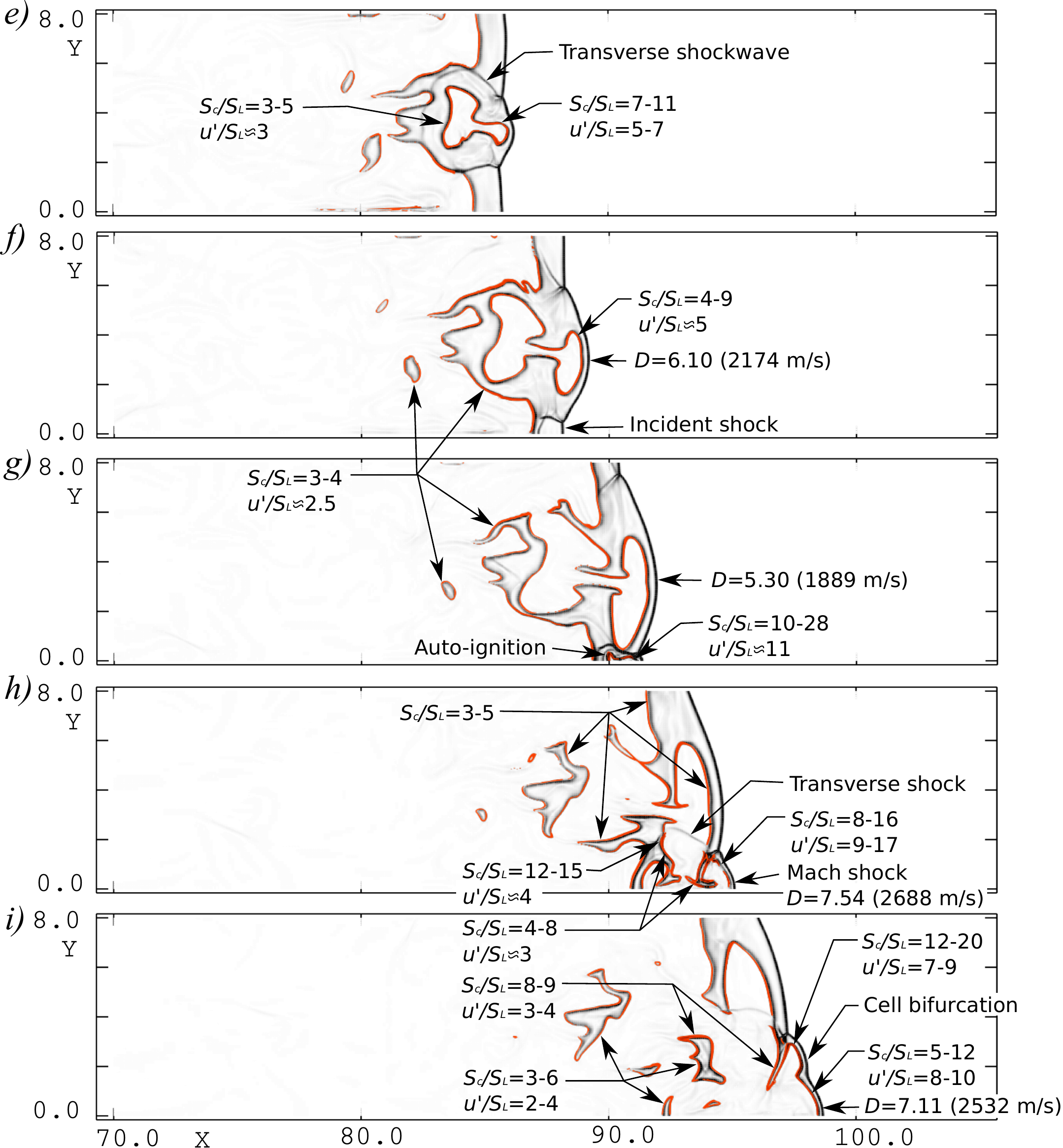}
	\caption{Onset of detonation initiation continued from Fig.\ \ref{fig.detinit}.}
	\label{fig.detinit2}
\end{figure}

In order to gain further understanding of the mechanisms which contribute to fast-flame acceleration and the onset of detonation, a detailed analysis of numerically obtained Schlieren flow fields \cite{Settles2001}, and corresponding reaction rates, was carried out for the final moments of DDT.  The simulation considered for the analysis, shown in Figs.\ \ref{fig.detinit} and \ref{fig.detinit2}, was for $C_\kappa=2.4$, as the measured $L_{DDT}$ was found to capture well that obtained from the corresponding experiment \cite{Ahmed2016}.  Although $L_{DDT}\approx150$ for this particular case, the very first instance where the local shock speed exceeded the CJ-detonation velocity was actually found to occur much sooner, around $\sim80\Delta_{1/2}$ downstream from the obstacles.  This discrepancy is noted by the fact that $L_{DDT}$ is a measure of when the average wave speed reaches the CJ-value, and not just a measure of the local phenomena.  Figure \ref{fig.detinit} indicates the exact location where the local fast-flame first accelerates to the point of detonation in a numerically obtained soot foil image.  In this case, the exact local event leading to DDT was found to occur at $(x,y)=(80,4)$.  The numerically obtained Schlieren images, in Figs.\ \ref{fig.detinit} and \ref{fig.detinit2}, show the density gradients present for several instances in time through the local flame acceleration process.  The figures also show the locations of instantaneous chemical reactions, $\overline{\dot{\omega}}(x,y)$, which are superimposed onto the Schlieren images in red.

In Fig.\ \ref{fig.detinit}, frame (\textit{a}), an incident shock wave was observed with a turbulent deflagration wave present in its wake.  While the shock was fairly planar, the deflagration wave contained large scale instabilities on its flame surface.  The source of the turbulent motions observed on the flame surface originated from the obstacles placed at $x=0$, which were used to quench the initial detonation wave.  Of particular interest are the magnitudes of the local flame consumption speeds and turbulence intensities, which were found to be greater in the unstable regions of the flame, as indicated.  For the unstable regions of the flame front, which protruded toward the leading shock wave, $S_c=3-7S_L$ (50-115 m/s).  For the relatively planar section of the flame front, further within the wake of the shock wave, $S_c=2-4S_L$ (33-66 m/s).  The unstable region of the flame front was also found to have larger velocity fluctuations, which were on the order of $u'\approx3S_L$ (50 m/s) versus $u'\approx2.5S_L$ (41 m/s) in the more stable region.  The increased velocity fluctuations in the unstable regions of the flame, which corresponded to enhanced combustion rates, arose from hydrodynamic instabilities associated with the large scale turbulent motions present.  By frame (\textit{b}), it was observed that the energy released due to the enhanced burning in the unstable regions of the deflagration produced pressure waves, as indicated.  By frame (\textit{c}), two separate pressure waves were found to collide with the leading shock wave near $(x,y)=(80,4)$.  By frame (\textit{d}), the local shock amplification which resulted from these pressure waves caused a small amount of gas to auto-ignite.  The energy deposition and volumetric expansion resulting from the auto-ignition kernel was sufficient to drive a newly formed Mach shock close to the CJ-detonation velocity:  $D=6.36$ (2266 m/s).  Directly behind this Mach shock, local flame consumption speeds up to $S_c=16S_L$ (264 m/s) were observed, which were much higher than the local turbulent velocity fluctuation speeds, where $u'\approx4-6S_L$ (66-99 m/s).  By frame (\textit{e}), shown in Fig.\ \ref{fig.detinit2}, a transverse shock was observed, which originated from the local explosion event.  Also, the newly formed hot spot was observed to protrude toward the shock front, whose forward motion was initially driven by shock compression.  Eventually, however, the local turbulent flame consumption speeds, in this region, slowed to $S_c=7-11S_L$ (116-182 m/s), while the turbulent velocity fluctuations increased in magnitude to $u'\approx5-7S_L$ (82-115 m/s).  Downstream from the Mach shock, the back-end of the hot spot propagated outward through turbulent surface reactions.  In this region, $S_c=3-5S_L$ (50-82 m/s).  By frame (\textit{f}), the Mach shock speed slowed to $D=6.10$ (2174 m/s), below the CJ-value of $D_{CJ}=6.40$. By this point, the hot spot propagated into unbunred gas solely through turbulent surface reactions.  Close to the Mach shock, the local turbulent flame consumption speeds slowed to $S_c=4-9S_L$ (66-148 m/s), while the turbulent velocity fluctuations were comparable at $u'\approx5S_L$ (82 m/s).  By frame (\textit{g}), the shock speed further decayed to $D=5.3$ (1889 m/s).  For a shocked particle of gas at this particular shock strength, following the discussion in \cref{sec.disco1}, the distance to auto-ignition is $\Delta_i\approx30$ (75 mm).  Since the flame was located only one $\Delta_{1/2}$ (2.5 mm) from the Mach shock, it is thus confirmed that the growth of this combustion region, which started as a hot spot, was driven purely through turbulent instabilities and surface reactions.

In frame (\textit{g}), the reflection of the transverse shock wave at the wall, located at $y=0$, triggered another local explosion in the unburned fuel mixture, as indicated.  This new hot spot formed purely due to intense shock compression, which then propagated through a combination of turbulent burning and further shock ignition associated with the newly formed Mach shock ahead of the explosion zone.  Here, the local turbulent flame consumption speeds were observed as high as $S_c=28S_L$ (462 m/s), which were much higher than the flame consumption speeds of the previous local explosion event.  Although the flame velocity was still subsonic relative to the shocked and unburned gas, it's high velocity and energy release rates were sufficient to amplify the leading Mach shock through the shock wave amplification by coherent energy release (SWACER) mechanism \cite{Lee1978}.  Also, the turbulent velocity fluctuations were found to be on the order of $u'\approx11S_L$ (182 m/s), which were much lower than the local flame consumption speed.  This suggests that the initial shock compression, which resulted from the newly formed Mach shock, likely contributed to the increased fuel consumption rates as the flame propagated in the forward direction.  By frame (\textit{h}), the velocity of the newly formed Mach shock was found to be $D=7.54$, overdriven above the CJ-detonation velocity of $D_{CJ}=6.43$.  For this particular shock strength, $\Delta_i\approx0.1$ (0.5 mm) due to shock compression alone, which is compatible with observations in the figure, where $\Delta_i=0.2$.  By this moment in time, however, the forward motion of the flame front was also enhanced by forward jetting of burned products into unburned products, behind the Mach shock.  This type of forward jetting is believed to arise due to Kelvin-Helmholtz instability of the shear layers behind newly formed triple points, where the Mach shock, incident shock, and transverse shock all meet \cite{Massa2007,Mach2011,Bhattacharjee2013b}.  It was found that increased turbulence due to the forward jetting lead to velocity fluctuations which were on the order of the local flame consumption speeds.  In the region behind the newly formed Mach shock, $S_c=8-16S_L$ (132-264 m/s), while $u'=9-280S_L$ (148-151 m/s).  Furthermore, Landau-Darrieus instability coupled with the forward jetting may have given rise to further auto-ignition of reactants, in locations where intense heat could be generated from the large scale folding of flame surfaces \cite{Valiev2007}.  Thus, turbulent burning and shock compression are both compatible burning mechanisms within this location.  Eventually, by frame (\textit{i}), the Mach shock velocity began to slow, owing to volumetric expansion of the wave front.  Also turbulent velocity fluctuations and local burning velocities decreased in magnitude as the local wave velocity decreased.  Behind the Mach shock, in frame (\textit{i}), $S_c=5-12S_L$ (82-198 m/s), while $u'=8-10S_L$ (132-165 m/s).  Also observed here was a cell-bifurcation, which is a typical feature of unstable detonations such as methane-oxygen \cite{Radulescu2009,Mach2011}.  In particular, such bifurcations have been found to arise due to the aforementioned wall jetting effect \cite{Mach2011}.  In this case, a new cell was formed, with $S_c=12-20S_L$ (198-330 m/s), while $u'=7-9S_L$ (116-148 m/s) behind it's leading shock wave.  Owing to the large magnitudes of velocity fluctuations, and even higher magnitudes of local flame consumption speeds, burning was likely due to a combination of turbulence and shock compression.  As these Mach shocks eventually decay in velocity, the burning propagates entirely as turbulent flames until the next shock reflection, or cell bifurcation.

Elsewhere in the flow fields of frames (\textit{h}) and (\textit{i}), burning was found to occur entirely as turbulent surface reactions, which propagated away from the local explosion events, and also into pockets of unburned fuel.  In general these turbulent flame consumption speeds are on the order of $S_c=3-8S_L$ (50-132 m/s), with $u'\approx3S_L$ (50 m/s).  Immediately behind the reflected transverse shock wave, however, local flame speeds were enhanced due to Richtmyer–Meshkov instability, with $S_c=12-15S_L$ (198-248 m/s).  The enhanced burning at these fuel interfaces, however, dampened quickly due to decoupling of velocity and thermal gradients, which is typical behaviour for high activation energy mixtures \cite{Massa2007}.

Although the average wave front did not accelerate to the CJ-detonation velocity until further downstream in the channel, near $x=150$, these local explosion events were key to generating local shock speeds periodically above the CJ-detonation velocity, which thus contributed to the overall acceleration of the wave front.  These explosion events generated strong transverse waves within the channel, which ultimately contributed to self-sustained detonation propagation, as observed in recent DDT experiments of Li et al. \cite{Li2017}.  Eventually, these periodic explosion events allowed for the wave to reach a self-sustained velocity, as observed by the velocity measurements in Fig.\ \ref{fig.detspeeds}.  Real detonations waves can thus be thought of as having a complex shock structure with trailing turbulent flames, which are coupled to periodic local explosions that give rise to an average propagation speed near the CJ detonation value.  It is noted, however, that during the propagation phase, much larger cells are expected than those observed in Fig.\ \ref{fig.detinit}.  In previous simulations of unobstructed detonation propagation in methane-oxygen \cite{Maxwell2016b}, cells widths were on the order of 10-20$\Delta_{1/2}$.  In the current investigation, however, the cells were much smaller; on the order of 5$\Delta_{1/2}$ (12.5 mm).  Also, it is noted that the cell size for this the particular mixture under investigation, according to Shepherd's detonation database \cite{Kaneshige1997}, should be ${\lambda}=16.5-19.0$ (41-47 mm).  Previously, Maxwell et al. \cite{Maxwell2016b} found that a value of $C_\kappa=6.7$ was required in order for detonations to re-produce the correct cell size during propagation.  This suggests that $C_\kappa$ takes on different values for the deflagration acceleration, and detonation propagation phases.  Accordingly, the authors here believe that the propagation phase contains much larger eddies, and is much more turbulent than the flame acceleration phase.

\section{Conclusions}

In the this study, DDT of methane-oxygen at low pressures was investigated using a state-of-the-art LES technique in order to address the wide range of turbulent scales present in highly compressible and reactive flows.  The method adopted here was the CLEM-LES method, which was previously used to simulated unobstructed detonation propagation, also in methane-oxygen at low pressures \cite{Maxwell2016b}.

It was found that turbulent velocity fluctuations and surface reactions play a major role which contribute to DDT events in such unstable mixtures.  As a precursor to DDT, it was confirmed that combustion waves must travel, on average, at velocities above the CJ-deflagration speed.  This limit was previously suggested by Ahmed et al. \cite{Saif2016} through experimental observation.  Furthermore, it was found that near the CJ-deflagration speed, shock compression alone cannot possibly describe the ignition in the wake of the leading shock wave.  Instead diffusive burning is necessary to maintain combustion wave velocities above the CJ-deflagration threshold.  It was also found that averaged turbulent velocity fluctuation magnitudes must be greater than the post-shock laminar flame speed in order to maintain combustion wave speeds above this CJ-deflagration limit.  Should velocity fluctuation magnitudes exceed this value, it was found that large scale instabilities are able to form on the early flame front development.  This leads to increased reaction rates, which generate pressure waves.  Eventually, these pressure waves coalesce and locally amplify the leading shock wave sufficiently to trigger the formation of a hot spot through shock compression.  This hot spot, or localized explosion, ultimately burns out through turbulent mechanisms.  The rapid burning of such hot spots, enhanced through surface reactions subject to turbulent instabilities, was found sufficient to drive transverse shock waves outward.  Upon reflection of such transverse waves with a wall, or other transverse waves, new local hot spots or local explosions subsequently form through shock compression, which again burn out through turbulent instabilities.  It was found that these local explosions are key to accelerating the wave to the self-sustaining CJ-detonation speed.  In this sense, the final stages of DDT behaves much like the propagation phase.  Previously, real detonations in unstable mixtures were found to travel below CJ-detonation speeds at most locations, and burning was generally found to occur as turbulent surface reactions.  Such detonations are only able to sustain their average CJ-detonation speeds through intense explosions at triple point collisions \cite{Maxwell2016b}.  This form of DDT also resembles that from shock-flame interaction experiments of Thomas et al. \cite{Thomas2001}.  In these past experiments, which involve ethylene-oxygen, shocked flame surfaces were found to amplify the leading shock through energy release associated with enhanced turbulent combustion.  The strengthened shock waves were then able to trigger auto-ignition hot spots, which thus contributed to detonation initiation.

A combustion regime diagram was also constructed for the simulations conducted here.  A key finding for the cases where DDT occurred was that flame surfaces in the flow field were found to lie predominantly in the thin-reaction zones regime.  In fact, some locations were also found to propagate as laminar flames and also broken reaction zones.  Owing to the large range of combustion regimes present in the flow field, this work confirms that simplified turbulent combustion models, such as flamelet and well stirred reactor models, cannot be used to model DDT of moderately unstable methane-oxygen mixtures.  This finding also justifies the use, and demonstrates the advantage of using the CLEM-LES to model such types of flow fields.  Finally, it was found that the $C_\kappa$ parameter was directly proportional to the turbulent length scales, or size of eddies, present in the flow field.  It was also found, however, that unlike previous simulations of detonation propagation \cite{Maxwell2016b}, increasing $C_\kappa$, or the turbulent length scale, did not necessarily increase the amount of turbulent fluctuations present.  As a result, a value of $C_\kappa=2.4$ was found to capture the correct DDT behaviour observed in experiments \cite{Ahmed2016}.  For the subsequent detonation propagation, however, this value was found to produce incorrect cellular behaviour.  Previously, a value of $C_\kappa=6.7$ was recommended for modelling detonations in methane-oxygen during the propagation phase \cite{Maxwell2016b}.  Thus, once detonation occurs, the turbulent length scales and mixing rates become much larger than during the flame acceleration phase leading up to detonation.  For future work, a dynamic procedure for obtaining $C_\kappa$ in compressible flows \cite{Chai2012} is recommended.

\section{Acknowledgements}

Funding for this work was supported by Shell and the Natural Sciences and Engineering Research Council of Canada through a Collaborative Research Grant.  The hard work of S. Ahmed and L. Maley are also acknowledged for their experimental contributions referenced in this work. Finally, S. Falle is also acknowledged for providing the numerical framework used in this work.


\section*{References}

\bibliography{keylatex}

\begin{thebibliography}{61}
\expandafter\ifx\csname natexlab\endcsname\relax\def\natexlab#1{#1}\fi
\providecommand{\url}[1]{\texttt{#1}}
\providecommand{\href}[2]{#2}
\providecommand{\path}[1]{#1}
\providecommand{\DOIprefix}{doi:}
\providecommand{\ArXivprefix}{arXiv:}
\providecommand{\URLprefix}{URL: }
\providecommand{\Pubmedprefix}{pmid:}
\providecommand{\doi}[1]{\href{http://dx.doi.org/#1}{\path{#1}}}
\providecommand{\Pubmed}[1]{\href{pmid:#1}{\path{#1}}}
\providecommand{\bibinfo}[2]{#2}
\ifx\xfnm\relax \def\xfnm[#1]{\unskip,\space#1}\fi
\bibitem[{{Buncefield Major Incident Investigation Board}(2008)}]{BMIB2008}
\bibinfo{author}{{Buncefield Major Incident Investigation Board}},
  \bibinfo{title}{{The Buncefield Incident 11 December 2005}},
  \bibinfo{type}{Final Report} \bibinfo{number}{Volumes 1 and 2}, Buncefield
  Major Incident Investigation Board, \bibinfo{year}{2008}.
\bibitem[{Johnson(2010)}]{Johnson2010}
\bibinfo{author}{M.~Johnson},
\newblock \bibinfo{title}{The potential for vapour cloud explosions - lessons
  from the {Buncefield} accident},
\newblock \bibinfo{journal}{J. Loss Prevent. Proc. Ind.} \bibinfo{volume}{23}
  (\bibinfo{year}{2010}) \bibinfo{pages}{921--927}.
\bibitem[{Pandey and Debnath(2016)}]{Pandey2016}
\bibinfo{author}{K.~M. Pandey}, \bibinfo{author}{P.~Debnath},
\newblock \bibinfo{title}{Review on recent advances in pulse detonation
  engines},
\newblock \bibinfo{journal}{Journal of Combustion} \bibinfo{volume}{2016}
  (\bibinfo{year}{2016}) \bibinfo{pages}{Article ID 4193034, 16 pages}.
\bibitem[{Maley(2015)}]{Maley2015}
\bibinfo{author}{L.~Maley}, \bibinfo{title}{On Shock Reflections in Fast
  Flames}, Master's thesis, University of Ottawa, \bibinfo{address}{Ottawa,
  Canada}, \bibinfo{year}{2015}.
\bibitem[{Ahmed(2016)}]{Ahmed2016}
\bibinfo{author}{M.~S.~A. Ahmed}, \bibinfo{title}{Run-Up Distance from
  Deflagration to Detonation in Fast Flames}, Master's thesis, Ottawa-Carleton
  Institute for Mechanical and Aerospace Engineering,
  \bibinfo{address}{University of Ottawa, Ottawa, Canada},
  \bibinfo{year}{2016}.
\bibitem[{Saif et~al.(2017)Saif, Wang, Pekalski, Levin, and
  Radulescu}]{Saif2016}
\bibinfo{author}{M.~Saif}, \bibinfo{author}{W.~Wang},
  \bibinfo{author}{A.~Pekalski}, \bibinfo{author}{M.~Levin},
  \bibinfo{author}{M.~I. Radulescu},
\newblock \bibinfo{title}{{Chapman-Jouguet} deflagrations and their transition
  to detonation},
\newblock \bibinfo{journal}{Proc. Combust. Inst.} \bibinfo{volume}{36}
  (\bibinfo{year}{2017}) \bibinfo{pages}{2771--2779}.
\bibitem[{Radulescu and Maxwell(2011)}]{Radulescu2011}
\bibinfo{author}{M.~I. Radulescu}, \bibinfo{author}{B.~M. Maxwell},
\newblock \bibinfo{title}{The mechanism of detonation attenuation by a porous
  medium and its subsequent re-initiation},
\newblock \bibinfo{journal}{J. Fluid Mech.} \bibinfo{volume}{667}
  (\bibinfo{year}{2011}) \bibinfo{pages}{96--134}.
\bibitem[{Zhu et~al.(2007)Zhu, Chao, and Lee}]{Zhu2007}
\bibinfo{author}{Y.~J. Zhu}, \bibinfo{author}{J.~Chao},
  \bibinfo{author}{J.~H.~S. Lee},
\newblock \bibinfo{title}{An experimental investigation of the propagation
  mechanism of critical deflagration waves that lead to the onset of
  detonation},
\newblock \bibinfo{journal}{Proc. Combust. Inst.} \bibinfo{volume}{31}
  (\bibinfo{year}{2007}) \bibinfo{pages}{2455--2462}.
\bibitem[{Lyamin et~al.(1991)Lyamin, Mitrofanov, Pinaev, and
  Subbotin}]{Lyamin1991}
\bibinfo{author}{G.~A. Lyamin}, \bibinfo{author}{V.~V. Mitrofanov},
  \bibinfo{author}{A.~V. Pinaev}, \bibinfo{author}{V.~A. Subbotin},
\newblock \bibinfo{title}{Propagation of gas explosion in channels with uneven
  walls and in porous media},
\newblock in: \bibinfo{editor}{A.~Borisov} (Ed.), \bibinfo{booktitle}{Dynamic
  Structure of Detonation in Gaseous and Dispersed Media}, volume
  \bibinfo{volume}{153}, \bibinfo{publisher}{kluwer}, \bibinfo{year}{1991}, pp.
  \bibinfo{pages}{51--75}.
\bibitem[{Teodorczyk et~al.(1989)Teodorczyk, Lee, and
  Knystautas}]{Teodorczyk1988}
\bibinfo{author}{A.~Teodorczyk}, \bibinfo{author}{J.~H.~S. Lee},
  \bibinfo{author}{K.~Knystautas},
\newblock \bibinfo{title}{Propagation mechanism of quasi-detonations},
\newblock \bibinfo{journal}{Symp. (Int.) Combust.} \bibinfo{volume}{22}
  (\bibinfo{year}{1989}) \bibinfo{pages}{1723--1731}.
\bibitem[{Arienti and Shepherd(2005)}]{Arienti2005}
\bibinfo{author}{M.~Arienti}, \bibinfo{author}{J.~E. Shepherd},
\newblock \bibinfo{title}{A numerical study of detonation diffraction},
\newblock \bibinfo{journal}{J. Fluid Mech.} \bibinfo{volume}{529}
  (\bibinfo{year}{2005}) \bibinfo{pages}{117--146}.
\bibitem[{Eckett et~al.(2000)Eckett, Quirk, and Shepherd}]{Eckett2000}
\bibinfo{author}{C.~A. Eckett}, \bibinfo{author}{J.~J. Quirk},
  \bibinfo{author}{J.~E. Shepherd},
\newblock \bibinfo{title}{The role of unsteadiness in direct initiation of
  gaseous detonations},
\newblock \bibinfo{journal}{J. Fluid Mech.} \bibinfo{volume}{421}
  (\bibinfo{year}{2000}) \bibinfo{pages}{147--183}.
\bibitem[{Radulescu and Maxwell(2010)}]{Radulescu2010}
\bibinfo{author}{M.~I. Radulescu}, \bibinfo{author}{B.~M. Maxwell},
\newblock \bibinfo{title}{Critical ignition in rapidly expanding self-similar
  flows},
\newblock \bibinfo{journal}{Phys. Fluids} \bibinfo{volume}{22}
  (\bibinfo{year}{2010}).
\bibitem[{Soloukhin(1969)}]{Soloukhin1969}
\bibinfo{author}{R.~I. Soloukhin},
\newblock \bibinfo{title}{Nonstationary phenomena in gaseous detonation},
\newblock \bibinfo{journal}{Proc. Combust. Inst.} \bibinfo{volume}{12}
  (\bibinfo{year}{1969}) \bibinfo{pages}{799}.
\bibitem[{Lundstrom and Oppenheim(1969)}]{Lundstrom1969}
\bibinfo{author}{E.~A. Lundstrom}, \bibinfo{author}{A.~K. Oppenheim},
\newblock \bibinfo{title}{On the influence of non-steadiness on the thickness
  of the detonation wave},
\newblock \bibinfo{journal}{Proc. R. Soc. Lond. A} \bibinfo{volume}{310}
  (\bibinfo{year}{1969}) \bibinfo{pages}{463--478}.
\bibitem[{Maxwell and Radulescu(2011)}]{Maxwell2011a}
\bibinfo{author}{B.~M. Maxwell}, \bibinfo{author}{M.~I. Radulescu},
\newblock \bibinfo{title}{Ignition limits of rapidly expanding diffusion
  layers: Application to unsteady hydrogen jets},
\newblock \bibinfo{journal}{Combust. Flame} \bibinfo{volume}{158 (10)}
  (\bibinfo{year}{2011}) \bibinfo{pages}{1946--1959}.
\bibitem[{Radulescu et~al.(2005)Radulescu, Sharpe, Lee, Kiyanda, Higgins, and
  Hanson}]{Radulescu2005}
\bibinfo{author}{M.~I. Radulescu}, \bibinfo{author}{G.~J. Sharpe},
  \bibinfo{author}{J.~H.~S. Lee}, \bibinfo{author}{C.~B. Kiyanda},
  \bibinfo{author}{A.~J. Higgins}, \bibinfo{author}{R.~K. Hanson},
\newblock \bibinfo{title}{The ignition mechanism in irregular structure gaseous
  detonations},
\newblock \bibinfo{journal}{Proc. Combust. Inst.} \bibinfo{volume}{30}
  (\bibinfo{year}{2005}) \bibinfo{pages}{1859--1867}.
\bibitem[{Makris et~al.(1993)Makris, Kamel, Kilambi, Knystautas, and
  Lee}]{Makris1993b}
\bibinfo{author}{A.~Makris}, \bibinfo{author}{A.~P.~M. Kamel},
  \bibinfo{author}{J.~Kilambi}, \bibinfo{author}{R.~Knystautas},
  \bibinfo{author}{J.~H.~S. Lee},
\newblock \bibinfo{title}{Mechanisms of detonation propagation in a porous
  medium},
\newblock in: \bibinfo{editor}{A.~L. Kuhl}, \bibinfo{editor}{J.~C. Leyer},
  \bibinfo{editor}{A.~A. Borisov}, \bibinfo{editor}{W.~A. Sirignano} (Eds.),
  \bibinfo{booktitle}{Dynamic Aspects of Detonation},
  \bibinfo{publisher}{American Institute of Aeronautics and Astronautics},
  \bibinfo{year}{1993}, pp. \bibinfo{pages}{363--380}.
\bibitem[{Makris et~al.(1995)Makris, Shafique, Lee, and
  Knystautas}]{Makris1995}
\bibinfo{author}{A.~Makris}, \bibinfo{author}{H.~Shafique},
  \bibinfo{author}{J.~H.~S. Lee}, \bibinfo{author}{R.~Knystautas},
\newblock \bibinfo{title}{Influence of mixture sensitivity and pore-size on
  detonation velocities in porous-media},
\newblock \bibinfo{journal}{Shock Waves} \bibinfo{volume}{5}
  (\bibinfo{year}{1995}) \bibinfo{pages}{89--95}.
\bibitem[{Slungaard et~al.(2003)Slungaard, Engebretsen, and
  Sonju}]{Slungaard2003}
\bibinfo{author}{T.~Slungaard}, \bibinfo{author}{T.~Engebretsen},
  \bibinfo{author}{O.~K. Sonju},
\newblock \bibinfo{title}{The influence of detonation cell size and regularity
  on the propagation of gaseous detonations in granular materials},
\newblock \bibinfo{journal}{Shock Waves} \bibinfo{volume}{12}
  (\bibinfo{year}{2003}) \bibinfo{pages}{301--308}.
\bibitem[{Fickett and Davis(1979)}]{Fickett1979}
\bibinfo{author}{W.~Fickett}, \bibinfo{author}{W.~C. Davis},
  \bibinfo{title}{Detonation Theory and Experiment},
  \bibinfo{publisher}{Dover}, \bibinfo{year}{1979}.
\bibitem[{Gamezo et~al.(2008)Gamezo, Ogawa, and Oran}]{Gamezo2008}
\bibinfo{author}{V.~N. Gamezo}, \bibinfo{author}{T.~Ogawa},
  \bibinfo{author}{E.~S. Oran},
\newblock \bibinfo{title}{Flame acceleration and {DDT} in channels with
  obstacles: Effect of obstacle spacing},
\newblock \bibinfo{journal}{Combust. Flame} \bibinfo{volume}{155}
  (\bibinfo{year}{2008}) \bibinfo{pages}{302--315}.
\bibitem[{Ogawa et~al.(2013)Ogawa, Oran, and Gamezo}]{Ogawa2013}
\bibinfo{author}{T.~Ogawa}, \bibinfo{author}{E.~S. Oran},
  \bibinfo{author}{V.~N. Gamezo},
\newblock \bibinfo{title}{Numerical study on flame acceleration and {DDT} in an
  inclined array of cylinders using an {AMR} technique},
\newblock \bibinfo{journal}{Computers Fluids} \bibinfo{volume}{85}
  (\bibinfo{year}{2013}) \bibinfo{pages}{63--70}.
\bibitem[{Houim et~al.(2016)Houim, Ozgen, and Oran}]{Houim2016}
\bibinfo{author}{R.~W. Houim}, \bibinfo{author}{A.~Ozgen},
  \bibinfo{author}{E.~S. Oran},
\newblock \bibinfo{title}{The role of spontaneous waves in the
  deflagration-to-detonation transition in submillimetre channels},
\newblock \bibinfo{journal}{Combust. Theor. Model.} \bibinfo{volume}{20 (6)}
  (\bibinfo{year}{2016}) \bibinfo{pages}{1068--1087}.
\bibitem[{Poludnenko(2017)}]{Poludnenko2017}
\bibinfo{author}{A.~Y. Poludnenko},
\newblock \bibinfo{title}{On {Chapman Jouguet} deflagrations},
\newblock \bibinfo{journal}{26th International Colloquium on the Dynamics of
  Explosions and Reactive Systems}  (\bibinfo{year}{2017})
  \bibinfo{pages}{paper 1247}.
\bibitem[{Johansen and Ciccarelli(2013)}]{Johansen2013}
\bibinfo{author}{C.~Johansen}, \bibinfo{author}{G.~Ciccarelli},
\newblock \bibinfo{title}{Modeling the initial flame acceleration in an
  obstructed channel using large eddy simulation},
\newblock \bibinfo{journal}{J. Loss Prevent. Proc. Ind.} \bibinfo{volume}{26}
  (\bibinfo{year}{2013}) \bibinfo{pages}{571--585}.
\bibitem[{Gaathaug et~al.(2012)Gaathaug, Vaagsaether, and
  Bjerketvedt}]{Gaathaug2012}
\bibinfo{author}{A.~V. Gaathaug}, \bibinfo{author}{K.~Vaagsaether},
  \bibinfo{author}{D.~Bjerketvedt},
\newblock \bibinfo{title}{Experimental and numerical investigation of {DDT} in
  hydrogen-air behind a single obstacle},
\newblock \bibinfo{journal}{Int. J. Hydrogen Energy} \bibinfo{volume}{37}
  (\bibinfo{year}{2012}) \bibinfo{pages}{17606--17615}.
\bibitem[{Yu and Navarro-Martinez(2015)}]{Yu2015}
\bibinfo{author}{S.~Yu}, \bibinfo{author}{S.~Navarro-Martinez},
\newblock \bibinfo{title}{Modelling of deflagration to detonation transition
  using flame thickening},
\newblock \bibinfo{journal}{Proc. Combust. Inst.} \bibinfo{volume}{35}
  (\bibinfo{year}{2015}) \bibinfo{pages}{1955--1961}.
\bibitem[{Emami et~al.(2015)Emami, Mazaheri, Shamooni, and
  Mahmoudi}]{Emami2015}
\bibinfo{author}{S.~Emami}, \bibinfo{author}{K.~Mazaheri},
  \bibinfo{author}{A.~Shamooni}, \bibinfo{author}{Y.~Mahmoudi},
\newblock \bibinfo{title}{{LES} of flame acceleration and {DDT} in
  hydrogen–air mixture using artificially thickened flame approach and
  detailed chemical kinetics},
\newblock \bibinfo{journal}{Int. J. Hydrogen Energy} \bibinfo{volume}{40 (23)}
  (\bibinfo{year}{2015}) \bibinfo{pages}{7395--7408}.
\bibitem[{Radulescu et~al.(2013)Radulescu, Sharpe, and Bradley}]{Radulescu2013}
\bibinfo{author}{M.~Radulescu}, \bibinfo{author}{G.~Sharpe},
  \bibinfo{author}{D.~Bradley},
\newblock \bibinfo{title}{A universal parameter quantifying explosion hazards,
  detonability and hot spot formation: The $\chi$ number},
\newblock \bibinfo{journal}{Proceedings of the Seventh Intational Seminar on
  Fire and Explosion Hazards}  (\bibinfo{year}{2013})
  \bibinfo{pages}{617--626}.
\bibitem[{Maxwell(2016)}]{Maxwell2016}
\bibinfo{author}{B.~M. Maxwell}, \bibinfo{title}{Turbulent Combustion Modelling
  of Fast-Flames and Detonations Using Compressible {LEM-LES}}, Ph.D. thesis,
  Ottawa-Carleton Institute for Mechanical and Aerospace Engineering,
  \bibinfo{address}{University of Ottawa, Ottawa, Canada},
  \bibinfo{year}{2016}.
\bibitem[{Menon and Kerstein(2011)}]{Menon2011}
\bibinfo{author}{S.~Menon}, \bibinfo{author}{A.~R. Kerstein},
\newblock \bibinfo{title}{The linear-eddy model},
\newblock in: \bibinfo{editor}{T.~Echekki}, \bibinfo{editor}{E.~I. Mastorakos}
  (Eds.), \bibinfo{booktitle}{Turbulent Combustion Modeling: Advances, New
  Trends and Perspectives}, \bibinfo{publisher}{Springer},
  \bibinfo{year}{2011}, pp. \bibinfo{pages}{221--247}.
\bibitem[{Maxwell et~al.(2017)Maxwell, Bhattacharjee, Lau-Chapdelaine, Falle,
  Sharpe, and Radulescu}]{Maxwell2016b}
\bibinfo{author}{B.~M. Maxwell}, \bibinfo{author}{R.~R. Bhattacharjee},
  \bibinfo{author}{S.~S.~M. Lau-Chapdelaine}, \bibinfo{author}{S.~A. E.~G.
  Falle}, \bibinfo{author}{G.~J. Sharpe}, \bibinfo{author}{M.~I. Radulescu},
\newblock \bibinfo{title}{Influence of turbulent fluctuations on detonation
  propagation},
\newblock \bibinfo{journal}{J. Fluid Mech.} \bibinfo{volume}{818}
  (\bibinfo{year}{2017}) \bibinfo{pages}{646--696}.
\bibitem[{Settles(2001)}]{Settles2001}
\bibinfo{author}{G.~S. Settles}, \bibinfo{title}{Schlieren and Shadowgraph
  Techniques}, \bibinfo{publisher}{Springer-Verlag}, \bibinfo{year}{2001}.
\bibitem[{Dennis et~al.(2014)Dennis, Maley, Liang, and Radulescu}]{Dennis2014}
\bibinfo{author}{K.~Dennis}, \bibinfo{author}{L.~Maley},
  \bibinfo{author}{Z.~Liang}, \bibinfo{author}{M.~I. Radulescu},
\newblock \bibinfo{title}{Implementation of large scale shadowgraphy in
  hydrogen explosion phenomena},
\newblock \bibinfo{journal}{Int. J. Hydrogen Energy} \bibinfo{volume}{39 (21)}
  (\bibinfo{year}{2014}) \bibinfo{pages}{11346--11353}.
\bibitem[{Radulescu et~al.(2007)Radulescu, Sharpe, Law, and
  Lee}]{Radulescu2007b}
\bibinfo{author}{M.~I. Radulescu}, \bibinfo{author}{G.~J. Sharpe},
  \bibinfo{author}{C.~K. Law}, \bibinfo{author}{J.~H.~S. Lee},
\newblock \bibinfo{title}{The hydrodynamic structure of unstable cellular
  detonations},
\newblock \bibinfo{journal}{J. Fluid Mech.} \bibinfo{volume}{580}
  (\bibinfo{year}{2007}) \bibinfo{pages}{31--81}.
\bibitem[{Kaneshige and Shepherd(1997)}]{Kaneshige1997}
\bibinfo{author}{M.~Kaneshige}, \bibinfo{author}{J.~Shepherd},
  \bibinfo{title}{Detonation database}, \bibinfo{type}{GALCIT Report}
  \bibinfo{number}{FM97-8}, California Institute of Technology: Aeronautics and
  Mechanical Engineering, \bibinfo{year}{1997}.
\bibitem[{Radulescu et~al.(2013)Radulescu, Wang, Saif, Levin, and
  Pekalski}]{Radulescu2015}
\bibinfo{author}{M.~I. Radulescu}, \bibinfo{author}{W.~Wang},
  \bibinfo{author}{M.~Saif}, \bibinfo{author}{M.~Levin},
  \bibinfo{author}{A.~Pekalski},
\newblock \bibinfo{title}{On {Chapman Jouguet} deflagrations},
\newblock \bibinfo{journal}{25th International Colloquium on the Dynamics of
  Explosions and Reactive Systems}  (\bibinfo{year}{2013})
  \bibinfo{pages}{paper 266}.
\bibitem[{Falle(1991)}]{Falle1991}
\bibinfo{author}{S.~A. E.~G. Falle},
\newblock \bibinfo{title}{Self-similar jets},
\newblock \bibinfo{journal}{Mon. Not. R. Astron. Soc.} \bibinfo{volume}{250}
  (\bibinfo{year}{1991}) \bibinfo{pages}{581--596}.
\bibitem[{Falle and Giddings(1993)}]{Falle1993}
\bibinfo{author}{S.~A. E.~G. Falle}, \bibinfo{author}{J.~R. Giddings},
  \bibinfo{title}{Numerical Methods for Fluid Dynamics IV},
  \bibinfo{publisher}{Oxford University Press}, \bibinfo{year}{1993}, pp.
  \bibinfo{pages}{337--343}.
\bibitem[{Maxwell et~al.(2015)Maxwell, Falle, Sharpe, and
  Radulescu}]{Maxwell2015}
\bibinfo{author}{B.~M. Maxwell}, \bibinfo{author}{S.~A. E.~G. Falle},
  \bibinfo{author}{G.~Sharpe}, \bibinfo{author}{M.~I. Radulescu},
\newblock \bibinfo{title}{A compressible-{LEM} turbulent combustion subgrid
  model for assessing gaseous explosion hazards},
\newblock \bibinfo{journal}{J. Loss Prevent. Proc. Ind.} \bibinfo{volume}{36}
  (\bibinfo{year}{2015}) \bibinfo{pages}{460--470}.
\bibitem[{Goodwin et~al.(2016)Goodwin, Moffat, and Speth}]{Goodwin2013}
\bibinfo{author}{D.~G. Goodwin}, \bibinfo{author}{H.~K. Moffat},
  \bibinfo{author}{R.~L. Speth}, \bibinfo{title}{Cantera: An object-oriented
  software toolkit for chemical kinetics, thermodynamics, and transport
  processes}, \bibinfo{howpublished}{{http://www.cantera.org}},
  \bibinfo{year}{2016}. \bibinfo{note}{Version 2.2.1}.
\bibitem[{Smith et~al.(2016)Smith, Golden, Frenklach, Moriarty, Eiteneer,
  Goldenberg, Bowman, Hanson, Song, Gardiner, Lissianski, and Qin}]{Smith2013}
\bibinfo{author}{G.~P. Smith}, \bibinfo{author}{D.~M. Golden},
  \bibinfo{author}{M.~Frenklach}, \bibinfo{author}{N.~W. Moriarty},
  \bibinfo{author}{B.~E. Eiteneer}, \bibinfo{author}{M.~Goldenberg},
  \bibinfo{author}{C.~T. Bowman}, \bibinfo{author}{R.~K. Hanson},
  \bibinfo{author}{S.~Song}, \bibinfo{author}{W.~C. Gardiner},
  \bibinfo{author}{V.~V. Lissianski}, \bibinfo{author}{Z.~Qin},
  \bibinfo{title}{{GRI-Mech 3.0}},
  \bibinfo{howpublished}{{http://www.me.berkeley.edu/gri\_mech/}},
  \bibinfo{year}{2016}.
\bibitem[{{O'Brien} et~al.(2017){O'Brien}, Towery, Hamlington, Ihme,
  Poludnenko, and Urzay}]{Obrien2017}
\bibinfo{author}{J.~{O'Brien}}, \bibinfo{author}{C.~A.~Z. Towery},
  \bibinfo{author}{P.~E. Hamlington}, \bibinfo{author}{M.~Ihme},
  \bibinfo{author}{A.~Y. Poludnenko}, \bibinfo{author}{J.~Urzay},
\newblock \bibinfo{title}{The cross-scale physical-space transfer of kinetic
  energy in turbulent premixed flames},
\newblock \bibinfo{journal}{Proc. Combust. Inst.} \bibinfo{volume}{36 (2)}
  (\bibinfo{year}{2017}) \bibinfo{pages}{1967--1975}.
\bibitem[{{Abdel-Gayed} et~al.(1984){Abdel-Gayed}, {Al-Khishali}, and
  Bradley}]{AbdelGayed1984}
\bibinfo{author}{R.~G. {Abdel-Gayed}}, \bibinfo{author}{K.~J. {Al-Khishali}},
  \bibinfo{author}{D.~Bradley},
\newblock \bibinfo{title}{Turbulent burning velocities and flame straining in
  explosions},
\newblock \bibinfo{journal}{Proc. R. Soc. Lond. A} \bibinfo{volume}{391}
  (\bibinfo{year}{1984}) \bibinfo{pages}{393--414}.
\bibitem[{Peters(1999)}]{Peters1999}
\bibinfo{author}{N.~Peters},
\newblock \bibinfo{title}{The turbulent burning velocity for large-scale and
  small-scale turbulence},
\newblock \bibinfo{journal}{J. Fluid Mech.} \bibinfo{volume}{384}
  (\bibinfo{year}{1999}) \bibinfo{pages}{107--132}.
\bibitem[{Tamadonfar and G{\"u}lder(2014)}]{Tamadonfar2014}
\bibinfo{author}{P.~Tamadonfar}, \bibinfo{author}{O.~L. G{\"u}lder},
\newblock \bibinfo{title}{Flame brush characteristics and burning velocities of
  premixed turbulent methane/air {Bunsen} flames},
\newblock \bibinfo{journal}{Combust. Flame} \bibinfo{volume}{161}
  (\bibinfo{year}{2014}) \bibinfo{pages}{3154--3165}.
\bibitem[{Dahoe et~al.(2013)Dahoe, Skjold, Roekaerts, Pasman, Eckhoff,
  Hanjalic, and Donze}]{Dahoe2013}
\bibinfo{author}{A.~E. Dahoe}, \bibinfo{author}{T.~Skjold},
  \bibinfo{author}{D.~J. E.~M. Roekaerts}, \bibinfo{author}{H.~J. Pasman},
  \bibinfo{author}{R.~K. Eckhoff}, \bibinfo{author}{K.~Hanjalic},
  \bibinfo{author}{M.~Donze},
\newblock \bibinfo{title}{On the application of the levenberg–marquardt
  method in conjunction with an explicit runge-kutta and an implicit rosenbrock
  method to assess burning velocities from confined deflagrations},
\newblock \bibinfo{journal}{Flow Turbul. Combust.} \bibinfo{volume}{91}
  (\bibinfo{year}{2013}) \bibinfo{pages}{281--317}.
\bibitem[{Fru et~al.(2011)Fru, Th\'{e}venin, and Janiga}]{Fru2011}
\bibinfo{author}{G.~Fru}, \bibinfo{author}{D.~Th\'{e}venin},
  \bibinfo{author}{G.~Janiga},
\newblock \bibinfo{title}{Impact of turbulence intensity and equivalence ratio
  on the burning rate of premixed methane–air flames},
\newblock \bibinfo{journal}{Energies} \bibinfo{volume}{4}
  (\bibinfo{year}{2011}) \bibinfo{pages}{878--893}.
\bibitem[{Short and Quirk(1997)}]{Short1997}
\bibinfo{author}{M.~Short}, \bibinfo{author}{J.~J. Quirk},
\newblock \bibinfo{title}{On the nonlinear stability and detonability limit of
  a detonation wave for a model three-step chain-branching reaction},
\newblock \bibinfo{journal}{J. Fluid Mech.} \bibinfo{volume}{339}
  (\bibinfo{year}{1997}) \bibinfo{pages}{89--119}.
\bibitem[{Sharpe and Maflahi(2006)}]{Sharpe2006}
\bibinfo{author}{G.~J. Sharpe}, \bibinfo{author}{N.~Maflahi},
\newblock \bibinfo{title}{Homogeneous explosion and shock initiation for a
  three-step chain-branching reaction model},
\newblock \bibinfo{journal}{J. Fluid Mech.} \bibinfo{volume}{566}
  (\bibinfo{year}{2006}) \bibinfo{pages}{163--194}.
\bibitem[{Fickett et~al.(1971)Fickett, Jacobson, and Schott}]{Fickett1971}
\bibinfo{author}{W.~Fickett}, \bibinfo{author}{J.~D. Jacobson},
  \bibinfo{author}{G.~L. Schott},
\newblock \bibinfo{title}{Calculated pulsating one-dimensional detonations with
  induction-zone kinetics},
\newblock \bibinfo{journal}{AIAA Journal} \bibinfo{volume}{10}
  (\bibinfo{year}{1971}) \bibinfo{pages}{514--516}.
\bibitem[{Lee et~al.(1978)Lee, Knystautas, and Yoshikawa}]{Lee1978}
\bibinfo{author}{J.~H. Lee}, \bibinfo{author}{R.~Knystautas},
  \bibinfo{author}{N.~Yoshikawa},
\newblock \bibinfo{title}{Photochemical initiation of gaseous detonations},
\newblock \bibinfo{journal}{Acta Astronautica} \bibinfo{volume}{5}
  (\bibinfo{year}{1978}) \bibinfo{pages}{971--982}.
\bibitem[{Massa et~al.(2007)Massa, Austin, and Jackson}]{Massa2007}
\bibinfo{author}{L.~Massa}, \bibinfo{author}{J.~M. Austin},
  \bibinfo{author}{T.~L. Jackson},
\newblock \bibinfo{title}{Triple-point shear layers in gaseous detonation
  waves},
\newblock \bibinfo{journal}{J. Fluid Mech.} \bibinfo{volume}{586}
  (\bibinfo{year}{2007}) \bibinfo{pages}{205--248}.
\bibitem[{Mach and Radulescu(2011)}]{Mach2011}
\bibinfo{author}{P.~Mach}, \bibinfo{author}{M.~I. Radulescu},
\newblock \bibinfo{title}{Mach reflection bifurcations as a mechanism of cell
  multiplication in gaseous detonations},
\newblock \bibinfo{journal}{Proc. Comb. Inst.} \bibinfo{volume}{33}
  (\bibinfo{year}{2011}) \bibinfo{pages}{2279--2285}.
\bibitem[{Bhattacharjee(2013)}]{Bhattacharjee2013b}
\bibinfo{author}{R.~R. Bhattacharjee}, \bibinfo{title}{Experimental
  Investigation of Detonation Re-initiation Mechanisms Following a {Mach}
  Reflection of a Quenched Detonation}, Master's thesis, University of Ottawa,
  \bibinfo{address}{Ottawa, Canada}, \bibinfo{year}{2013}.
\bibitem[{Valiev(2007)}]{Valiev2007}
\bibinfo{author}{D.~Valiev}, \bibinfo{title}{The role of {Landau-Darrieus}
  instability in flame dynamics and deflagration-to-detonation transition},
  Ph.D. thesis, KTH Royal Institute of Technology, \bibinfo{address}{Stockholm,
  Sweden}, \bibinfo{year}{2007}.
\bibitem[{Radulescu et~al.(2009)Radulescu, Papi, Quirk, Mach, and
  Maxwell}]{Radulescu2009}
\bibinfo{author}{M.~I. Radulescu}, \bibinfo{author}{A.~Papi},
  \bibinfo{author}{J.~J. Quirk}, \bibinfo{author}{P.~Mach},
  \bibinfo{author}{B.~M. Maxwell},
\newblock \bibinfo{title}{The origin of shock bifurcations in cellular
  detonations},
\newblock \bibinfo{journal}{22nd International Colloquium on the Dynamics of
  Explosions and Reactive Systems}  (\bibinfo{year}{2009})
  \bibinfo{pages}{paper 106}.
\bibitem[{Li et~al.(2017)Li, Li, Nguyen, Teo, Chang, and Khoo}]{Li2017}
\bibinfo{author}{L.~Li}, \bibinfo{author}{J.-M. Li}, \bibinfo{author}{V.~B.
  Nguyen}, \bibinfo{author}{C.~J. Teo}, \bibinfo{author}{P.-H. Chang},
  \bibinfo{author}{B.~C. Khoo},
\newblock \bibinfo{title}{A study of detonation re-initiation through multiple
  reflections in a 90-degree bifurcation channel},
\newblock \bibinfo{journal}{Combust. Flame} \bibinfo{volume}{180}
  (\bibinfo{year}{2017}) \bibinfo{pages}{207--216}.
\bibitem[{Thomas et~al.(2001)Thomas, Bambrey, and Brown}]{Thomas2001}
\bibinfo{author}{G.~Thomas}, \bibinfo{author}{R.~Bambrey},
  \bibinfo{author}{C.~Brown},
\newblock \bibinfo{title}{Experimental observations of flame acceleration and
  transition to detonation following shock-flame interaction},
\newblock \bibinfo{journal}{Combust. Theor. Model.} \bibinfo{volume}{5:4}
  (\bibinfo{year}{2001}) \bibinfo{pages}{573--594}.
\bibitem[{Chai and Mahesh(2012)}]{Chai2012}
\bibinfo{author}{X.~Chai}, \bibinfo{author}{K.~Mahesh},
\newblock \bibinfo{title}{Dynamic k-equation model for large-eddy simulation of
  compressible flows},
\newblock \bibinfo{journal}{J. Fluid Mech.} \bibinfo{volume}{699}
  (\bibinfo{year}{2012}) \bibinfo{pages}{385--413}.

\end{thebibliography}

\end{document}